\documentstyle[emulateapj]{article}
\newcommand{\ew}{\ifmmode{W_{\lambda}} \else $W_{\lambda}$\fi}
\newcommand{\ax}{\ifmmode{\alpha_x} \else $\alpha_x$\fi} 
\newcommand{\aox}{\ifmmode{\alpha_{ox}} \else $\alpha_{ox}$\fi} 
\newcommand{\kms}{\ifmmode~{\rm km~s}^{-1}\else ~km~s$^{-1}~$\fi}
\newcommand{\ros}{{\em ROSAT{\rm}}}

\received{August 9, 2000}
\accepted{}

\lefthead{Forster, Green, Aldcroft, Vestergaard, Foltz, \& Hewett}
\righthead{Emission Line Properties of the Large Bright Quasar Survey}

\begin{document}
\title{Emission Line Properties of the Large Bright Quasar Survey}

\author{Karl Forster,}
\affil{email: {\em kforster@cfa.harvard.edu}}

\author{Paul J. Green, Thomas L. Aldcroft,}
\affil{Harvard-Smithsonian Center for Astrophysics, 60 Garden St.,
Cambridge, MA 02138} 
\affil{email: {\em pgreen@cfa.harvard.edu, aldcroft@cfa.harvard.edu}}

\author{Marianne Vestergaard}
\affil{The Ohio State University, 140 West 18th Avenue, 
Columbus, OH 43210-1173} 
\affil{email: {\em vester@astronomy.ohio-state.edu}}

\author{Craig B. Foltz}
\affil{Multiple Mirror Telescope Observatory, University of
Arizona, Tucson, AZ 85721} 
\affil{email: {\em cfoltz@as.arizona.edu}}

\and

\author{Paul C. Hewett}
\affil{Institute of Astronomy, Madingley Road, Cambridge, CB3 0HA, UK} 
\affil{email: {\em phewett@ast.cam.ac.uk}}

\begin{abstract}
We present measurements of the optical/UV emission lines for a large
homogeneous sample of 993 quasars from the Large Bright Quasar
Survey. Our largely automated technique accounts for continuum breaks
and galactic reddening, and we perform multicomponent fits to emission
line profiles, including the effects of blended iron emission, and of
absorption lines both galactic and intrinsic.  Here we describe the
fitting algorithm and present the results of line fits to the LBQS
sample, including upper limits to line equivalent widths when
warranted.  The distribution of measured line parameters, principally
\ew\, and FWHM, are detailed for a variety of lines, including upper
limits.  We thus initiate a large-scale investigation of correlations
between the high energy continuum and emission lines in quasars, to be
extended to complementary samples using similar techniques.
High quality, reproducible measurements of emission lines for 
uniformly selected samples will advance our understanding of active
galaxies, especially in a new era of large surveys selected by a variety
of complementary methods.
\end{abstract}


\section{Introduction}
\label{intro}

Quasars, some of the most luminous objects in the Universe, allow us
to observe towards the start of cosmic time, probing extremes of
physics, and illuminating  the intervening matter.  Though most of  
the QSOs now known were discovered by their unique spectral energy
distributions (SEDs) or emission lines, several interrelated and
fundamental questions remain:    

\smallskip

 1)  What significant correlations exist between the observed SED 
and emission lines in large uniform quasar samples?  

 2)  How are the observed SED and emission lines affected by intrinsic
 absorption and orientation?  

 3)  Can the shape of the intrinsic SED be determined? 

 4)  What effect does the intrinsic SED have on emission lines
and what does that tell us about physical conditions in the QSO?

\smallskip

The production of emission lines in QSO spectra is widely attributed
to photoionization and heating of the emitting gas by the UV to X-ray
continuum (e.g., Ferland \& Shields 1985; Krolik \& Kallman 1988).
Emission lines from a given ion are particularly sensitive to photons
of energy above the corresponding ionization threshold, but ionization
from excited states and heating via free-free and H$^-$ absorption
also determine the line's principal ionizing/heating continuum.
Many important lines respond to the extreme ultraviolet (EUV) or soft
X-ray continuum.  Unfortunately, the EUV band is severely obscured by
Galactic absorption, although some constraints on the EUV ionizing
continuum are available through analysis of the adjacent UV and soft
X-ray windows.

The overall similarity of QSO emission line spectra has on occasion
been taken as evidence of fairly uniform physical conditions in the
broad emission line region (BELR).  At first, this encouraged the
assumption that clouds in the BELR inhabit a narrow swath of parameter
space in density, size, and ionization parameter ($U$).  Observed
correlations of line equivalent widths (hereafter, \ew) with SEDs
present challenges to geometric and photoionization models that need
to be answered.  Photoionization pioneers such as Mushotzky \& Ferland
(1984) ran models on a single cloud, while photoionization models
assuming a distribution of clouds (i.e., not a single $U$) predict a
strong dependence of $W_{\lambda}$ on the (assumed power law) slope of
the SED (Binette et al. 1989; Baldwin et al. 1995; Korista, Baldwin,
\& Ferland 1998).

Large uniform samples of QSO emission line measurements are key to
test photoionization models, but are also necessary for other
important tests of quasar triggering and evolution.  Tests of the
binary vs. lens hypothesis for quasar pairs at the same redshift
require quantitative measures of the probability of finding the
measured spectral similarity derived from complete samples (Mortlock,
Webster \& Francis 1999).  Since optically-selected QSOs at similar
redshift may tend to be more similar than 2 QSOs chosen entirely at
random, studies of QSO spectral properties should be based on samples
large enough to provide statistically significant subsamples across a
wide range of luminosity and redshift (Peng et al. 1999).

In a small, complete sample of optically-selected QSOs studied by Laor
et al. (1997), the strongest correlation found between X-ray continuum
and optical emission line parameters was of the soft X-ray spectral
slope (\ax), and the FWHM of H$\beta$.  Strong correlations between \ax~
and the relative strengths of [\ion{O}{3}] $\lambda$5007 and iron
emission were seen both there and in previous studies, particularly
Boroson \& Green (1992) (hereafter BG92). A principal component
analysis (PCA) was employed by BG92 to examine an unwieldy set of
emission line and continuum correlations within a sample of 87 QSOs
taken from the Palomar-Green Bright Quasar Survey (BQS; Schmidt
\& Green 1983). This indicated that the strongest sources of variance
in the optical spectra, the ``primary eigenvector'', involved an
anti-correlation between the strength of [\ion{O}{3}] $\lambda$5007
and iron emission that was also linked to the FWHM of H$\beta$. This
work was extended by Marziani et al. (1996) (hereafter M96) and
Sulentic et al. (2000) to include a study of the spectral region
around \ion{C}{4} $\lambda$1549. These papers identified Narrow-line
type 1 Seyfert galaxies (NLS1) and steep spectrum radio galaxies as
occupying opposite ends of the primary eigenvector.  Other work
utilizing the high spectral resolution of the {\em Hubble Space
Telescope}~ Faint Object Spectrograph ({\em HST} FOS) showed a link
between \ion{Al}{3} $\lambda$1859, \ion{Si}{3}] $\lambda$1892, and
\ion{C}{3}] $\lambda$1909 emission line strengths and the primary
eigenvector but suffered from the small sample size and a focus on
NLS1 or the low redshift radio quiet objects in the BQS (Aoki \&
Yoshida 1999; Kuraszkiewicz et al. 2000).

The samples of AGN used by most authors are almost entirely taken from
BG92 and so do not represent a statistically complete sample and may
not be truly representative of the QSO population (Wisotzki et
al. 2000). They also deal with the strongest emission features in the
OUV spectrum and, while several hypotheses have been proposed, the
physical origin of the observed correlations remains a puzzle.  There
is also a concern that biases within these small samples may have
exaggerated the significance of the primary eigenvector.

Even in the clear presence of extrinsic effects such as absorption
along the line of sight, emission lines have been used to infer the
strength and shape of the high energy SED.  As an example, the
similarity of UV emission-line properties in broad absorption line
(BAL) and non-BAL QSOs (Weymann et al. 1991) has been cited as
evidence that orientation is the cause of the BAL phenomenon (i.e.,
that {\em all} radio-quiet QSOs have BAL clouds).  However, BALQSOs
are now known to exhibit markedly weak X-ray emission as a class
(Green \& Mathur 1996; Gallagher et al. 1999; Brandt, Laor, \& Wills
2000).  But if similar emission  
lines indeed vouch for similar intrinsic high energy SEDs, then the
large X-ray to optical continuum flux ratio (\aox) values observed for
BALQSOs are likely to be caused by strong absorption along the
line-of-sight rather than by differences in their intrinsic SEDs.  In
fact, the soft X-ray deficit requires absorption at least an order of
magnitude larger than those estimated from optical/UV (OUV) spectra
alone and the UV and X-ray absorbers have yet to be definitively
identified as one (e.g., see discussion in Mathur, Wilkes, \& Elvis
1998).  Since BALQSOs may be heavily absorbed, the question of whether
similar emission lines are testimony for similar intrinsic SEDs must
be answered through the systematic study of line and continuum
correlations in {\em unabsorbed} QSOs.

Even in supposedly unabsorbed QSOs, a strong inverse correlation
exists between optical iron emission, and \ion{O}{3} $\lambda$ 5007 in QSOs
(BG92).  Iron emission is under-predicted by a large factor in
standard photoionization models relative to observations (Collin \&
Joly 2000), and the continuum source responsible for creating
rest-frame optical iron emission lines comes almost entirely from
X-rays above about 1keV.  QSOs with large \aox\, of which BALQSOs
represent an extreme, have not only weak 
\ion{O}{3}, but feeble narrow line emission in general (Green 1996).  We
suggest that the observed correlations may be an effect of absorption
of the intrinsic high energy continuum by thick clouds interior to the
NELR.  The X-ray heating of these clouds makes them emit copious
\ion{Fe}{2}, lines, but prevents the ionizing continuum from reaching the
NELR.  This is reinforced by another observed correlation
- that the most highly absorbed BALQSOs, the low-ionization BALQSOs,
albeit X-ray quiet, are generally very strong optical
\ion{Fe}{2}, emitters (Stocke et al. 1992; Lipari, Terlevich, \& 
Macchetto 1993). Our interpretation needs to be tested across a range 
of absorption, in large representative QSO samples.

We have therefore undertaken a major study of quasar line emission,
with careful accounting for absorption lines and blended iron
emission, using largely automated procedures on carefully-selected
samples. The analysis of samples of QSOs with more than a few hundred
spectra requires some amount of automation to give consistent
results. The largest sample of QSO spectra currently available is that
of the Large Bright Quasar Survey and here we describe the initial
results from the measurement of this sample and the analysis methods
that will also be applied to other large samples of QSO spectra that
will become available in the near future.

\section{The LBQS Sample }
\label{lbqssample}

The Large Bright Quasar Survey (LBQS; see Hewett, Foltz \& Chaffee
1995) is a sample of 1058 QSOs selected using the Automatic Plate
Measuring Machine from UK Schmidt photographic direct and objective
prism plates.  The combination of quantifiable selection techniques,
including overall spectral shape, strong emission lines, and
redshifted absorption features has been shown to be highly efficient
at finding QSOs with $0.2<z<3.3$, a significantly broader range than
previous work.  The LBQS thus avoids selection effects common in other
optical quasar samples that tend to exclude weak-lined quasars or to
under-sample certain redshift ranges or colors.  There appears to be a
deficiency in the numbers of QSOs with $z \sim 0.8$ in the LBQS
sample, although the effect is not statistically significant (Hewett
et al. 1995).  The LBQS has been for the last decade the principal
source of intermediate brightness optically--selected quasars
available; 1/9 of the Veron-Cetty \& Veron (1998) catalog of QSOs and
$\sim50\%$ of all known quasars within a similar magnitude range.
Radio data are currently available for about 1/3 of the sample (Hooper
et al. 1995) and soft X-ray fluxes and upper limits for at least 85\%
of the sample can be obtained from the {\em ROSAT} All-Sky Survey.
Follow-up optical spectra with S/N$\approx$10 in the continuum at
4500\AA~ and $6~-~10$\AA\, resolution were obtained between 1986 and
1990 at the MMT on Mt Hopkins, Arizona and the 2.5m duPont in Las
Campanas, Chile for all LBQS QSOs.  Error spectra ($1\sigma$) are
available for 1009 of the objects.  The selection criteria and
observing procedures for each of the survey fields can be found in
Hewett et al. (1995) and references therein (LBQS Papers I-VI). Three
QSOs appear here that were identified subsequently to the 1055 QSOs in
the Hewett et al. (1995) LBQS catalog. They are 0052+0148 (z=0.595 ),
1027$-$0149 (z=0.754), and 2132$-$4227 (z=0.569), see Hewett, Foltz,
\& Chaffee (2001; in preparation) for further details.

\section{Analysis of the Spectra}

The LBQS spectra were analyzed using {\it Sherpa} \footnotemark, a
generalized fitting engine designed primarily for spatially resolved
spectroscopy of observations with NASA's {\it Chandra} X-ray
Observatory. {\it Sherpa} enables data to be modeled using a variety
of optimization methods and with a number of built-in statistical
tests. The ease with which user-models can be created and the ability
to simultaneously model a number of input data sets that may be in
different formats (e.g. modeling an ASCII and a PHA style FITS file
together) makes {\it Sherpa} suitable for our goals. In brief, the
process begins with modeling the continuum emission and then
accounting for any emission by iron complexes in the spectra. We then
model the emission lines using Gaussian profiles and, after a search
for significant absorption features, the results are inspected and the
emission line fitting procedure repeated to improve the model of each
spectrum.  The model parameters were determined from a minimization of
the $\chi^{2}$ statistic with modified calculations of uncertainties
in each bin (Gehrels 1986). We found that a Powell optimization method
gave a balance between an efficient processing time and consistency of
results.  Because of the large database of observations the modeling
of the spectra proceeds in a largely non-interactive manner.

\footnotetext{{\it Sherpa} is available from the {\it Chandra} science
center ({\tt http://asc.harvard.edu}).}

We do not present measurements of emission lines for objects with
strong BALs.  The diversity in the strength and shape of BALs
precludes even semi-automated measurements, and their emission line
properties have already been well-characterized for a strongly
overlapping sample in Weymann et al. (1991). We also exclude objects
that have multiple narrow absorption features that significantly mask
the true shape and strength of the emission features of Lyman $\alpha$
and \ion{C}{4} $\lambda$1549.  A list of the 65 objects excluded from
this study can be found in Table 1.  This list includes two objects
not previously classified as BAL in LBQS papers I-VI, 0059$-$2545 and
1242+1737.  The measurements for the remaining 993 LBQS spectra are
presented here.

\subsection{The continuum}

Our first step is to model the continuum emission by fitting one or
two power laws to ranges in the spectrum that are free from any strong
emission lines. We found the second power law component to be
necessary due to the steepening of the continuum towards the UV in low
redshift objects, i.e. for spectra that extended redward of rest frame
4200\AA.  A list of the continuum modeling `windows', along with the
nearest strong emission lines, can be found in Table 2.  We apply a
Galactic reddening correction to the power law continua that follows
the prescription given by Cardelli, Clayton, \& Mathis (1989) which
uses:

$$A (\lambda) = E (B - V) [ aR_{\rm V} + b ]$$

\noindent where $a$ and $b$ are polynomial functions of wavelength
derived for 3.3$\mu$m $\ge \lambda \ge 1000$ \AA. We use a ratio of
total to selective extinction of $R_{\rm V} = 3.1$, the standard value
for the diffuse ISM, and take the relationship between the color
excess and the line of sight column of neutral Hydrogen for each
object to be

$$E ( B - V ) = {{N_{\rm H} {\rm (Galactic)} (10^{20} {\rm cm}^{-2})}
\over {58.0}}$$

\noindent (Bohlin, Savage, \& Drake 1978).

Many of the spectra in the LBQS sample depart from a simple reddened
power law continuum for wavelengths blueward of $\sim$ 4000\AA~ in the
observed frame as noted in the original LBQS papers.  This depression
in the observed ultraviolet is due to the use of 2.5 arcsecond
circular apertures in the MMT observations coupled with the effects of
atmospheric dispersion at large zenith distances and a guider system
that was red-sensitive.  To model this part of the continuum, we use a
polynomial of up to 2nd order with the flux tied to that of the power
law continuum model in the nearest continuum window with $\lambda_{c}$
(observed frame) $> 4000$\AA.  We did not apply a polynomial continuum
to objects with $z \ge 2.42$ because the Lyman $\alpha$ emission line
appears redward of 4000\AA~ and the continuum on the blue side of
Lyman $\alpha$ can be significantly reduced by Lyman $\alpha$ forest
lines.

The polynomial continuum model is not physically meaningful and may be
inaccurate, particularly where strong iron emission complexes appear
around the \ion{Mg}{2} $\lambda$2800 emission line, i.e. for

$$ 2200 {\rm \AA} \lesssim \lambda_{\rm rest} \lesssim 3100 {\rm
\AA}$$

We investigated the effect of the polynomial continuum on \ew~
measurements and on the error estimates for the emission line
parameters by comparing lines of the same species that occurred in the
polynomial continuum region to those that did not. A
Kolmogorov-Smirnov test (e.g., Press et al. 1992) shows that the
distributions of \ew~ measurements for the UV iron emission,
\ion{Si}{4}+\ion{O}{4} $\lambda$1400, \ion{C}{4}, and \ion{Mg}{2}
emission lines are significantly different when measured above the
polynomial continuum compared to when measured using a power law
continuum. The median increase in \ew~ for \ion{C}{4} and \ion{Mg}{2}
is $\sim$ 20\% when these lines fall above the polynomial continuum
but genuine observational trends like the Baldwin Effect (Baldwin
1977; Osmer, Porter, \& Green 1994) may account for this measurement
trend.  The intrinsic continuum luminosity of objects with
\ion{C}{4} redshifted to $\lambda$ (observed frame) $> 4000$\AA~
is on average twice that of the lower redshift objects where
\ion{C}{4} falls above the polynomial continuum.  A 20\% increase in
\ew~ for \ion{C}{4} and \ion{Mg}{2} is predicted using the
relationships between \ew~ and continuum luminosity determined by
Zamorani et al. (1992) but may not fully account for the differences
seen in the measurements of \ew~ for these and other emission lines.
Rather than correcting for this effect, we added a systematic
uncertainty of 20\% to the estimation of emission line strengths and
upper limits for all emission lines measured above the polynomial
continuum.

Finally, we visually inspected the continuum model for each
observation, and found that minor adjustments were required in $\sim
20$\% of the sample; a large proportion of these were caused by strong
intrinsic absorption lines.

\subsection{Iron emission}

Our second step accounts for iron emission line complexes in the
spectra. This emission occurs most strongly in the regions around the
\ion{Mg}{2} $\lambda$2800 and H$\beta$ emission lines. We follow the
prescription used by BG92 of subtracting a template of iron emission
lines created from the spectrum of I Zw 1, an NLS1 which exhibits the
typically strong iron emission of this class of AGN.  We use two
templates: the optical emission line template is identical to that
used by BG92 and corrects for iron emission between 4400\AA~ $<
\lambda_{\rm rest} <$ 7000\AA, while the UV template was developed by
Vestergaard \& Wilkes (2000) from {\em HST} FOS observations of I~Zw~1
and covers the region 1250\AA~ $<
\lambda_{\rm rest} <$ 3100\AA. The FWHM of the iron emission features seen in
I~Zw~1 is similar to that of the H$\beta$ emission line (900 \kms) and
so sets the minimum template FWHM. We convolved this template with a
series of Gaussian functions of increasing width to produce a grid of
38 template spectra with 900 $\lesssim$ FWHM $\lesssim 10000$
\kms, conserving the total flux in each template. This provides us
with a nominal resolution of 250 km s$^{-1}$, more than adequate for
the quality of the spectra in the LBQS sample.

The first part in modeling the iron emission is to compare the 2000 km
s$^{-1}$ templates to windows in each spectrum on both sides of the
\ion{Mg}{2} $\lambda$2800 emission line and redward of \ion{C}{3}]
$\lambda$1909 for the UV emission, and to either side of the H$\beta$
+ [\ion{O}{3}] $\lambda\lambda$4959,5007 complex for the optical
emission (see Table 2). After the relative amplitude of the iron
emission has been modeled, the full grid of templates is applied to
the spectra to determine the approximate FWHM of the iron
emission. This is followed by a final modeling of the template
amplitude simultaneously with the FWHM of the template
spectra. Adjustments are made from a visual inspection of the results
to give the best representation of the iron emission complexes present
in each spectrum.

Note that we measure the strengths of the iron emission in the UV and
optical independently. This is important as the relative strength of
the UV and optical iron emission may allow an approximate
determination of the conditions in the emitting region, e.g. density
and $U$ (Verner et al. 1999).  UV and optical iron emission may also
correlate differently with X-ray emission.  Green et al. (1995) found
that QSOs in the LBQS with strong UV Fe\,II emission (based on the
iron feature under [Ne~IV]$\lambda 2423$) are anomalously X-ray bright
in the \ros\, passband.  By contrast, several studies (e.g., Corbin
1993, Lawrence et al. 1997) find an anti-correlation between soft
X-ray luminosity and $R(Fe\,II)$ (the equivalent width ratio of
optical iron emission to H$\beta$).  Several such intriguing
correlations have been noted, but among samples of varying size and
using heterogeneous measurement techniques.  Our project attempts to
remedy this situation using large samples and uniform analysis.

When first scaling the UV iron template in the LBQS spectra to the
iron emission complexes straddling \ion{Mg}{2} ($\sim \lambda \lambda
2200 - 3300$), we found that the template tended to over-predict the
iron emission line strength blueward of \ion{C}{4} $\lambda$1549. This
occurred in 27 of the 238 spectra that cover both \ion{C}{4} and
\ion{Mg}{2} and that have measurable iron emission.  The UV iron
template is based on an {\em HST} FOS spectrum of I~Zw~1 for which the
region blueward of \ion{C}{4} was observed six months earlier than the
remaining spectrum. I~Zw~1 clearly brightened between the two
observations, for which reason the bluest sub-spectrum was scaled to
match the redder spectrum.  The reasons for the over-prediction of
iron flux blueward of \ion{C}{4} in our sample when using the I~Zw~1
template are not clear, but may include a change in iron emission
strength and multiplet ratios with luminosity indicating variations in
the physical conditions to which the iron emission is sensitive
(e.g. Netzer 1980).  See Vestergaard \& Wilkes (2000) for further
details and discussion. To compensate for this effect, the UV iron
template flux in the region $\lambda_{\rm rest} < 1530$\AA~ was
reduced by 50\%, resulting in an improved agreement with all the
sample spectra.

\subsection{Emission lines}

The emission lines in the spectra are modeled by the addition of
Gaussian features at the expected wavelengths to the continuum + iron
template model. The initial value of the peak flux near the position
of the line is estimated from the raw data assuming a continuum level
under each emission line calculated by fitting a straight line between
the continuum windows nearest the line to be modeled.  The initial
estimate of the FWHM of the emission line component is assumed to be
3000 to 5000 km s$^{-1}$ for broad components, and 350 km s$^{-1}$ for
narrow components.  These reflect the typical widths of emission lines
produced in the BELR and NELR (e.g. Peterson 1997).  Where the S/N in
the spectra are high enough, the Lyman $\alpha$, \ion{C}{4},
\ion{Mg}{2}, and H$\beta$ emission lines may be modeled using two
Gaussian components.  The initial value of the peak amplitude of each
component is set to be 40\% of the peak flux estimated from the data.
Lower S/N spectra are modeled with a single Gaussian but a visual
inspection of the results is made and a second Gaussian component is
added where significant residuals appear.

The FWHM and peak amplitude of the Gaussian model components are then
optimized within a spectral region covering the emission line. The
position of the emission line is fixed at the expected wavelength in
the initial modeling until after the presence of narrow absorption
features has been determined (\S3.4). The region of the spectrum used
in modeling an emission feature is chosen to best optimize the
component of interest, e.g. the Lyman $\alpha$ emission line is
simultaneously fitted with \ion{N}{5} $\lambda$1240 and \ion{O}{1}
$\lambda$1305 because the broad component to Lyman $\alpha$ may have a
significant effect on the spectrum near \ion{O}{1}.  A list of the
emission lines modeled in the LBQS spectra is presented in Table
3. There are a number of important emission lines (e.g. \ion{C}{2}
$\lambda$1336) that are not modeled here due to the quality of the
spectra, but will be included in the application of this procedure to
samples of higher spectral resolution (e.g. {\em HST} FOS spectra).

\subsection{Absorption lines}

One of the goals of this study is to determine the frequency of
occurrence of absorption in the spectra of QSOs. We used the program
FINDSL (Aldcroft 1993) to detect significant narrow absorption
features and model them with multi-component Gaussian fitting. Our
input to FINDSL is the sum of the reddened continuum, iron template(s)
and all the emission line profiles from the first {\it Sherpa}
modeling. This constitutes a `continuum' from which FINDSL detects
significant deviations in the observed spectrum.  We rescale the error
array of each spectrum for use with FINDSL so that $\chi^{2}_{\nu}$
becomes unity, which yields a better estimate of the true error array
and aids in the detection of significant absorption features.  We
exclude the Lyman $\alpha$ forest region blueward of rest frame
1065\AA~ from the absorption line detection, and also exclude the
Balmer continuum region (between 3360\AA~ -- 3960\AA~) where the
global continuum model may lie below the spectrum causing many
spurious absorption lines to be detected.

The absorption line measurements from FINDSL were then included with
the emission line and continuum results for further modeling using
{\it Sherpa}. In this second iteration the emission line positions are
also fitted, and the absorption lines modeled simultaneously with the
emission line components. The results from this automated process were
inspected and adjustments made to those spectra that the algorithms
could not successfully model. This was not unexpected as the wide
variety of spectra in the sample and the chance superposition of
absorption features on the emission lines makes it unlikely that any
fully automatic procedure would be 100\% successful.

\subsection{Error analysis}

An estimate of the 2$\sigma$ error range for each parameter of the
continuum and emission line components was determined from the
$\chi^{2}$ confidence interval bounds ($\Delta\chi^{2}=4.0$).  Where
the amplitude of an emission line could not be constrained to within
2$\sigma$, the line position was reset to the expected wavelength and
the FWHM to a median value that was determined from the distribution
of well constrained emission line measurements within the LBQS sample.
These fixed emission line parameters are given in Table 3. The
emission line amplitude was then re-fitted and, if remaining
unconstrained at the 2$\sigma$ level, then the 2$\sigma$ upper limit
on the amplitude was estimated. The determination of upper limits for
lines that may be present at low flux levels in the spectra of QSOs is
one of the primary objectives of this work. This allows the use of
survival analysis techniques in determining a more realistic
distribution of emission line strengths.

\section{Emission Line results}
\label{lines}

The total number of emission lines measured from the spectra in the
LBQS sample are presented in Table 4. The numbers of upper limits
measured and the number of two component emission lines are also
tabulated. The full table of all emission line measurements and
uncertainties of the 993 spectra modeled here is available from the
ApJ in the electronic journal at {\tt
http://www.journals.uchicago.edu/ApJ/}, but due to the large size
cannot be reproduced in the printed journal. To show the form and
content of the large electronic table we present in Table 5 a digested
version containing measurements of the emission lines present in three
LBQS spectra.  The emission line measurements are quoted to a
significance level determined by the 2$\sigma$ uncertainties in the
parameters (e.g. Bevington \& Robinson 1992).  The format is as
follows: Each object is represented by a row containing the
designation and redshift followed by a row for each measured emission
line. The name of the emission line is given in column (1), column (2)
gives the FWHM in \kms, and column (3) gives the offset of the peak of
the Gaussian emission line model, in \kms, from the expected position
based on the tabulated redshift. Note that no peak offset measurements
appear for the iron emission line measurements. Column (4) gives the
rest frame equivalent width of the emission line in \AA. Each emission
line parameter is quoted with positive and negative 2$\sigma$ error
estimates. The errors quoted for \ew~ are based in the uncertainties
in the amplitude and FWHM of the Gaussian model and do not include an
error from an uncertainty in the underlying continuum flux level.  For
emission lines where only an upper limit on \ew~ could be measured,
there are no values quoted for the peak offset because we fixed the
position of the line at its expected wavelength. The FWHM value also
has no associated errors in such a case because it was fixed at an
approximate median value for the sample.  (See \S3.5.)  Note that some
detected but poorly constrained lines may have an error estimate for
the \ew~ measurement even while the FWHM is fixed (see \S3.5 and Table
4). Finally, column (5) gives the number of narrow absorption features
used in modeling the emission lines. They are tabulated in the row of
the closest emission line to the position of the absorption feature,
e.g. see Lyman $\alpha$ (broad) and \ion{O}{1} for LBQS 2354$-$0134 in
Table 5. The format of the electronic version of Table 5 is explained
in the Appendix.

To generate a measure of the strength of iron emission in each
spectrum, the integrated flux of the model template in the regions
2240\AA~ -- 2655\AA~ (blueward of \ion{Mg}{2}) or 4434\AA~ -- 4684\AA~
(blueward of H$\beta$) is used and combined with the continuum
specific flux at 2448\AA~ and 4559\AA~ respectively to calculate
\ew. For objects where these regions were not present in the observed
spectrum, we extrapolate the continuum to these positions. Note that
we integrate the optical iron template flux over a wavelength range
that matches the window used by BG92.

We present in Table 6 the continuum parameters for the same LBQS
spectra that appear in Table 5. The full table is available in
electronic form with one row for each QSO in the sample. The format of
Table 6 is as follows. Columns (1), (2) and (3) give the QSO
designation, redshift, and the Galactic neutral Hydrogen Column
density (in units of $10^{20}$ cm$^{-2}$) respectively. The latter
measurements are taken from the Bell Laboratory \ion{H}{1} survey
(Stark et al. 1992).  Columns (4) - (6) contain the measured values of
the power law continuum; continuum slope, the {\em rest frame}
wavelength at which the continuum is normalized, and the continuum
flux at that wavelength (ergs cm$^{-2}$ s$^{-1}$ \AA$^{-1}$)
respectively, along with their 2$\sigma$ uncertainties.  Note that in
Table 6 the normalization units are $10^{-16}$ but in the electronic
table the units are $10^{-14}$.  Column (7) lists the slope of the
second power law continuum if used, the normalization of which is
identical to the first power law continuum. Column (8) presents the
order of the polynomial continuum if required to model the blue region
of the spectra.  The format of the electronic version of Table 6 is
explained in the Appendix.

We present in Figures 3a, 3b, and 3c the profiles and residuals from
the models of the three spectra presented in Tables 5 and 6. The
measurements presented here were chosen to show a range of spectral
qualities present in the LBQS sample.  Fig. 3 includes a number of
panels for each spectrum in each case all wavelengths are observed
frame and fluxes are $10^{-14}$ ergs cm$^{-2}$ s$^{-1}$ \AA$^{-1}$.
The top panel has three sections, the upper section shows the
continuum model overlying the spectrum, including error bars on each
bin, the middle section displays the iron emission template plotted
over the spectrum (no error bars) and the lower section shows template
profile alone. Note that the flux scales are different in these
sections to display the template emission more clearly.

This panel is followed by a number of separate panels, one for each
emission line region.  These emission line panels have three sections;
the top section shows the total continuum + emission line model over
the spectrum. (The model will include flux from the iron emission line
template(s) if used.  For clarity, flux errors on each bin are not
shown in this section.)  The middle section shows the residuals from
the emission line model and include the error bars for each spectral
bin. The lower section shows the profiles of the emission line models,
with multiple components separated into single Gaussian profiles. The
shape of the iron emission template used in the modeling of the
spectrum is also shown in the lower panel. Dashed vertical lines
indicate the expected position of emission lines based on tabulated
redshifts.  Absorption line profiles are shown along the top edge of
the lower panels, e.g. the Lyman $\alpha$ region panel for LBQS
2354$-$0134 in Fig.3c.  For the panels that show the emission line
models of [\ion{Ne}{5}], [\ion{O}{2}], and [\ion{Ne}{3}] the top
sections also plot the local continua used for these lines (see
Appendix).  Note that to help visualize the quality of the emission
line models the wavelength scale is identical for each of the emission
line panels and within each panel the flux scale is identical in the
sections displaying the total model, residuals, and individual model
components.

\section{Overall Statistics}

The analysis of the LBQS sample of QSO spectra has provided us with
over 8000 emission line measurements, of which approximately 30\% are
upper limits to the equivalent width of low intensity lines. Without
the correct inclusion of these ``censored'' data in the analysis of
this complete sample, a biased estimation of the properties may occur
and a number of important relationships may be masked or, even worse,
appear more significant than in reality.

We use a non-parametric survival analysis technique to estimate the
means and medians that characterize the emission line parameter
distributions. The Kaplan-Meier (KM) estimator of a sample
distribution is a maximum likelihood method that provides a
reconstruction of information lost by censoring of data. This well
established statistical technique has been throughly examined for use
in astronomical studies by Feigelson \& Nelson (1985), Isobe,
Feigelson, \& Nelson (1986) and references therein. We recommend these
papers to the reader.

The astronomical survival statistics package ASURV Rev 1.1 (Isobe \&
Feigelson 1990; LaValley, Isobe \& Feigelson 1992) was used to provide
the means, errors on the mean, and the medians of the \ew~
distributions for the LBQS sample given in Table~7 and the
distributions of \ew~ shown in Figure 4. The format of Table~7 is as
follows. Column (1) lists the emission line or line blend
measured. The distribution for single Gaussian component models are
tabulated separately from narrow and broad components. The
distribution of sum of the \ew~ of the broad and narrow component
models included with the single component measurements is also
tabulated. Columns (2) to (4) give the number of detected emission
lines, their mean \ew~ and the standard deviation (SD) of the
distribution of \ew~.  Columns (5) and (6) list the total number of
emission lines measured for each species and the number of upper
limits estimated for \ew. Columns (7) and (8) give the KM means, error
on the means, and medians of \ew~ for each line. Columns (9) to (11)
give the number, mean and median of the FWHM of the Gaussian
components used to model each emission feature. All equivalent widths
are rest frame and the error on the \ew~ means is estimated from the
1$\sigma$ of the KM reconstructed distribution.  Note that there are
no upper limit measurements for two component emission line fits, in
cases where any of the parameters of the two components could not be
constrained to 2$\sigma$, a single Gaussian was used to model the
line.

We should add a note of caution here about the reconstructed
distribution of emission line properties for samples that contain a
large proportion of censored data (e.g. [\ion{Ne}{5}], \ion{He}{2}
$\lambda$4686). Formally, an emission line is detected if the
amplitude of a feature can be constrained to within 2$\sigma$, based
on the $\chi^{2}$ confidence interval bounds. The performance of the
KM estimator degrades if the censored fraction is high or if the
pattern of censored measurements is not random. We do not believe the
latter is a concern here because the observing procedure used in the
generation of the sample spectra produced spectra of similar continuum
S/N, independent of the strength of emission features and the
intrinsic luminosity of the QSO. In support of this conclusion
consider the statistics of the \ion{O}{1} emission line, where
approximately half of the spectra with coverage in the region of this
line possess detected emission features. Detected emission lines have
0.6 $\lesssim \ew
\lesssim$ 14.0 \AA~ but the range of upper limits for the undetected
lines is $0.2 \lesssim \ew \lesssim 15.0 $\AA~. The large proportion
of censored data for some emission lines lowers our confidence in the
KM estimate of the means and medians of a reconstructed \ew~
distribution. Any conclusions based on these emission lines should
rather use the results from the detected lines alone.

The histograms shown in Figure 4 are taken from the KM estimation of
the number of data points in each bin. The \ew~(\AA) and FWHM (\kms)
distribution for each emission line are shown in the upper and lower
panels for each emission line respectively.  For emission lines that
have been modeled using two Gaussian components, we include the
distribution for the narrow and the broad components separately from
the single component lines. For completeness we also show in Figure 4
the distribution of the full sample of measured lines where the \ew~
of broad+narrow components of an emission line have been summed and
included in the single component distribution (dashed line
histograms).

\section{Discussion}

To examine the robustness of the analysis procedure in OUV spectra
with improved S/N and spectral resolution we have modeled two
composite QSO spectra using the techniques described above. The first
was constructed by Francis et al. 1991 (hereafter F91) from 718
spectra in the LBQS sample itself, and the second from {\em HST} FOS
observations of 101 QSOs presented in Zheng et al. (1997) (hereafter
Z97). We find that even where the limit of two Gaussian components for
an emission line does not reproduce the line profile very well, the
measured \ew~ is within 10\% of values tabulated in the above
papers. The exception is for \ion{Mg}{2} where the F91 measurement is
higher and the Z97 measurement is lower than that measured using the
techniques presented above. Both discrepancies can be attributed to
the presence of iron emission features blending into the wings of the
\ion{Mg}{2} emission line. Although F91 fitted a global continuum and
corrected for iron emission using local spline fits to emission
complexes, the \ew~ was measured as the total flux above the continuum
and iron emission across more than 50\AA~ in the spectrum and so may
include some hidden iron emission. This is also likely to be the
explanation of a similar discrepancy in the measurements of H$\gamma$
+ [\ion{O}{3}] $\lambda$4363 and [\ion{O}{3}] $\lambda$4959 in the F91
composite.  The measurements in Z97 use a local continuum which is
difficult to determine accurately in regions where iron emission
complexes are strong, and may result in their lower \ew~ measurement.
This confirms that correct accounting for iron emission is crucial in
the measurement of emission lines in QSO spectra.

We can also compare the global properties of the LBQS emission line
measurements with the measurements of these composite QSO spectra.
The Z97 composite shows larger \ew~ in Lyman $\beta$ +
\ion{O}{6}, Lyman $\alpha$, and \ion{C}{4}.  The discrepancy 
remains even if the means of only our detected emission lines are
compared. (The inclusion of the censored data will naturally lower the
means of a distribution, but has little effect on the distribution of
strong lines like Lyman $\alpha$ and \ion{C}{4} which are well
constrained in nearly all cases).  We suspect that the difference is
attributable to the large (60\%) fraction of radio loud objects in the
heterogeneous sample used to construct the Z97 composite. The LBQS
sample is predominantly Radio Quiet (RQ) with $\sim$ 10\% Radio Loud
(RL) (Hooper et al. 1995) and Z97 showed that there are significant
differences in the strength of these lines between RL and RQ
populations, independent of luminosity. A marginally significant RL/RQ
difference was also seen in a sample of 255 of the optically brightest
QSOs in the LBQS sample (Francis, Hooper,
\& Impey 1993). A more detailed examination of this effect will be
made in a subsequent paper that will contrast the continuum and
emission line properties within the full LBQS sample.

In contrast, the \ew~ of the emission features in the F91 composite
are lower than the mean and medians of the measured emission lines in
the LBQS sample.  Although the F91 composite was created using a large
proportion of the same spectra measured here, it is not surprising
that the measurements should differ. Lyman\,$\alpha$, which shows the
largest dispersion of relative flux within the sample (see Fig. 5 in
F91), is most affected by narrow absorption features.  Allowance for
the effects of narrow absorption increases the \ew~ modeled with
Gaussian components relative to an estimate based on the integration
of flux above a continuum. For [\ion{Ne}{3}] and H$\delta$, the F91
composite indicates a stronger feature than suggested by the KM mean
presented in Table~7. However, the measurements are similar when only
the detected lines are studied.

There has been no investigation of inclusion of censored data on the
emission line and continuum correlations, particularly the Baldwin
effect (Baldwin 1977; Zamorani et al. 1992; Osmer, Porter, \& Green
1994) and the correlations associated with the primary eigenvector of
spectral variance in AGN (BG92, M96, Sulentic et al. 2000).  Previous
studies have focussed on a small number of bright emission features;
particularly \ion{C}{4} $\lambda$1549, \ion{He}{2} $\lambda$4686,
H$\beta$, \ion{O}{3} $\lambda$5007 and iron emission.  We briefly
summarize the published measurements for QSO samples of BG92 (87 low z
QSOs, optical spectra), McIntosh et al. (1999) (32 high z QSOs with IR
spectra), M96 (52 QSOs with optical and UV spectra) and Sulentic et
al. 2000 (125 QSOs) in Table~8. Where upper limits are quoted, we have
used survival analysis to enable a more direct comparison to our
results in Table~7. Some of the published work has also been presented
separately as RL or RQ subsamples.  While all these studies suffer
from differing selection effects, the comparison samples also differ
from the LBQS sample in redshift range and magnitude limits.

The most notable difference between the properties of the LBQS and the
measurements shown in Table~8 can be seen in the relative strength of
optical iron emission, the mean value of which appears significantly
higher in BG92 and MS96.  The region used to measure the \ew~ of
optical iron emission is identical to that used in BG92 (and
subsequent authors) and a test using the methods described in \S3 to
measure the BG92 sample spectra (kindly supplied by T. Boroson) shows
reasonable ($\sim$ 20\%) agreement in \ion{Fe}{2} \ew.  The
measurement of UV iron emission in M96 appears similar to the mean of
the distribution of iron emission strength measured in the LBQS
sample. However the window used by M96 to characterize the iron
emission strength was 1550 -- 1750\AA~ and the iron template flux in
this region is approximately 6.5\% of the flux in the region used for
the study of the LBQS sample presented here (i.e. the mean UV iron
emission in M96 is 15 times higher than the mean in the LBQS
sample!). The unusually strong iron emission was noted by M96 as a
possible bias in their sample.  The heterogeneous sample (52 AGN for
which $HST$ FOS spectra covering \ion{C}{4} were available in 1994 and
for which the authors had matching optical spectra) has an
overabundance of strong iron emitters relative to the BG92 sample. The
difference in iron emission distributions contrasts to the similarity
seen in the distributions of \ew~ for [\ion{O}{3}]$\lambda$5007. The
strongest correlations reported for optical spectra of AGN involve
iron emission, H$\beta$ and [\ion{O}{3}].  We expect that our new,
more homogeneous emission line measurements, of iron in particular,
may significantly reshape the discussion of correlations that produce
the largest variance in PCA analysis of quasar emission lines.

\section{Summary}

This paper makes available measurements from the largest database of
OUV spectra of QSOs to date. Our analysis of 993 spectra of similar
quality from the LBQS yields over 8000 measurements of the 20 most
prominent emission lines between rest frame 1025\AA~ -- 5900\AA~. We
have accounted for the effects of blended iron emission complexes,
narrow absorption lines, continuum breaks, and we include upper limits on
the strength of low intensity emission lines.  The measurements are
presented with uncertainties generated in an objective manner and with
a rigorous inclusion of censored data in tabulated emission line
parameter distributions.  A more exhaustive analysis of the
relationships between the emission line and continuum properties of
the LBQS sample will be presented in a later paper.

High quality, reproducible measurements of emission lines for 
uniformly selected samples will advance our understanding of active
galaxies, especially in a new era of large surveys selected by a variety
of complementary methods, such as the Sloan Digital Sky Survey
(SDSS; York et al. 2000), the FIRST Bright Quasar Survey (White et al.
2000), or the Chandra Multiwavelength Project (Green et al. 1999).

\bigskip

The authors gratefully acknowledge support provided by NASA through
grant NAG5-6410 (LTSA). CBF acknowledges the
support of NSF grant AST 98-03072. KF thanks
Matt Malkan and the UCLA Division of Astronomy \& Astrophysics for
their hospitality.

\clearpage

\centerline{APPENDIX}
\bigskip

\noindent{\it Notes on specific emission lines}

\noindent {\bf Lyman $\beta$ + \ion{O}{6} $\lambda$1035} --- A flat
`pseudo' continuum and a single Gaussian emission feature were used to
model the spectrum near this emission line blend.  The resulting
equivalent widths should be viewed as approximate due to the nature of
the QSO spectra in this region. Only in one spectrum of the sample was
the Lyman $\beta$ and \ion{O}{6} emission lines clearly distinct, but
the spectrum was modeled with a single Gaussian for consistency.

\bigskip

\noindent {\bf \ion{He}{2} $\lambda$1640} --- The single Gaussian
component used to model the region on the red side of \ion{C}{4}
$\lambda$1549 will account for emission from \ion{He}{2}
$\lambda$1640, [\ion{Ne}{5}] $\lambda\lambda$1575,1593, [\ion{Ne}{4}]
$\lambda\lambda$1602,1609, \ion{Si}{2} $\lambda$1650, [\ion{O}{3}]
$\lambda\lambda$1661,1663,1668, \& \ion{Al}{2} $\lambda$1670.  The use
of multiple Gaussian components in all but a few spectra did not
improve the model fit to this region due to the quality of the LBQS
sample. The tabulated values for this emission line component,
particularly the FWHM, will be much larger than expected for just the
\ion{He}{2} line.

\bigskip

\noindent {\bf [\ion{Ne}{5}] $\lambda$3426, [\ion{O}{2}]
$\lambda$3728, [\ion{Ne}{3}] $\lambda$3869} --- The global continuum
model was not successful for the region near these lines due to the
blend of Balmer emission lines from transitions with $m \ge 7$ in the
low resolution spectra and so a local power law continuum was created.
This was measured from windows 30\AA \hskip 0.1cm wide at least 30\AA
\hskip 0.1cm from the expected line position. The emission lines are
measured above this local continuum.

\bigskip

\noindent {\bf \ion{He}{2} $\lambda$4686} --- For spectra where two
Gaussian were used to model the H$\beta$ emission line, the H$\beta$ broad
component was included in the model fit of the \ion{He}{2}
$\lambda$4686 emission line.

\bigskip

\noindent{\it Individual Objects}

\noindent {\bf LBQS 1206+1052} --- This QSO was assigned a redshift of 
0.402 $\pm$ 0.005 (Hewett et al. 1995) based on the cross correlation
with the Francis et al. (1991) LBQS composite spectrum. Although all
the features in the spectrum will contribute to the redshift estimate
the stronger features will contribute more than the weak so it is
unclear as to the cause of the discrepancy as the strongest lines of
[\ion{O}{3}] $\lambda\lambda$ 4959,5007 are clearly not at the
tabulated redshift.  Only the measurement of the \ion{Mg}{2} emission
line agrees with this redshift, this is likely due to the presence of
an absorption feature on the blue side of \ion{Mg}{2} that is not
modeled in the automated fitting procedure used here. Other lines in
the spectrum all show a lower redshift. The value adopted here is $z =
0.396 \pm 0.003$ based on the median of 10 emission measurements and
the 1$\sigma$ value of the distribution of z.

\bigskip

\noindent {\bf LBQS 0023+0228} --- The spectrum of this QSO shows very
strong and narrow (FWHM $\simeq$ instrumental resolution) forbidden
emission lines, no evidence of broad H$\beta$ emission and only a weak
flat continuum.  The strength of [\ion{O}{2}] (\ew~ = 148
$^{+8}_{-11}$ \AA) and [\ion{O}{3}] (\ew~ = 258 $^{+12}_{-14}$ \AA) in
the spectrum of LBQS 0023+0228 is much higher than all the other QSOs
in the LBQS except for LBQS 0004+0224 which shows a rising blue
continuum and strong broad emission from \ion{Mg}{2} and H$\beta$ as
well as having forbidden line FWHM $\gtrsim$ 1200 km s$^{-1}$.  The
spectrum of LBQS 0023+0228 resembles that of a starburst galaxy
(e.g. M82) rather than a QSO. The presence of these two strong narrow
line objects in the sample does not affect the parameter distributions
due to the large numbers of objects in each sample.

\bigskip

\noindent{\it Electronic Tables}

The electronic version of Table 5 has an identical format to that
published here but includes one row for each emission line that is
listed in Table 3 even if the line lies outside the wavelength range
of the observed spectrum. The electronic table contains zeros for
unmeasured emission line parameters, rather than the (...)  present in
printed Table 5. This will aid in the use of this large machine
readable ASCII table.  The first line of the electronic table contains
shortened column headings. This is followed by 31 lines for each QSO
(a total of 30753 lines). The first line for each object tabulates the
QSO designation and redshift and has a format ({\tt a10,f6.3}). This
is followed by 30 rows with a format ({\tt a16,6i7,3f9.2,i3}), note
that the positive and negative error estimates appear in {\em
separate} columns.  Where only an upper limit for \ew~ of an expected
emission line is measured, the value is given in the positive error
column of \ew. The FWHM for these lines is given with zeros in the
error columns (the position offset is by definition 0~\kms).  Note
that there are emission lines where the FWHM and position offset could
not be constrained but the \ew~ is constrained to 2$\sigma$.

The format of the electronic version of Table 6 is identical to that
presented here and as with the electronic version of Table 5, the
positive and negative errors have separate columns. The first line of
the electronic table contains the column headings and is followed by a
single row for each object with the ASCII format ({\tt
a10,2f6.3,3f7.2,i6,3f11.5,3f7.2,i3}).  The normalization units for the
power law continuum (columns 8, 9, and 10 in the electronic table) are
10$^{-14}$ erg cm$^{-2}$ s$^{-1}$ \AA$^{-1}$.  We stress that in the
electronic form of Tables 5 and 6 zeros appear in all blank spaces
that would normally appear blank.

\clearpage


\begin{deluxetable}{lcclclccl}
\tablewidth{350pt}
\small
\tablenum{1}
\tablecaption{LBQS QSOs excluded from emission line measurements}
\tablehead{
Designation & Redshift & Ref\tablenotemark{a} & Notes  & \quad & Designation & 
Redshift & Ref\tablenotemark{a} & Notes
}
\startdata
0004$+$0147  & 1.710 & II  & &&  1230$+$1627B & 2.735 & i   & 1,2 \nl
0010$-$0012  & 2.154 & II  & &&      1230$+$1705  & 1.420 & III & \nl      
0013$-$0029  & 2.083 &     & 1,2 && 1231$+$1320  & 2.380 & I  \nl
0018$-$0220  & 2.596 &     & 1 && 1235$+$1453  & 2.699 & I   & \nl
0018$+$0047  & 1.835 & II  & &&  1235$+$0857  & 2.898 & I   & \nl
0019$+$0107  & 2.130 & II  & &&  1235$+$1807B & 0.449 & I   & \nl
0021$-$0213  & 2.293 & II  & &&  1239$+$0955  & 2.013 & I   & \nl
0022$+$0150  & 2.826 & II  & &&  1240$+$1516  & 2.297 & i   & 1,2 \nl
0025$-$0151  & 2.076 & II  & &&  1240$+$1551  & 0.573 & iii & 5 \nl
0029$+$0017  & 2.253 & II  & &&  1240$+$1607  & 2.360 & III & \nl
0045$-$2606  & 1.242 & V   & &&  1242$+$1737  & 1.863 &     & 1,2,3,4 \nl
0051$-$0019  & 1.713 & IV  & &&  1243$+$0121  & 2.796 & III & \nl
0054$+$0200  & 1.872 & IV  & &&  1314$+$0116  & 2.686 & III & \nl
0059$-$2545  & 1.955 &     & 3,4 && 1331$-$0108  & 1.881 & III & \nl
0059$-$0206  & 1.321 & IV  & &&  1332$-$0045  & 0.672 &     & 6 \nl
0059$-$2735  & 1.593 & V   & &&  1333$+$0133  & 1.577 & iii & 2 \nl
0100$-$2809  & 1.768 & V   & &&  1442$-$0011  & 2.226 & III & \nl
0103$-$2753  & 0.848 & V   & &&  1443$+$0141  & 2.451 & III & \nl
0109$-$0128  & 1.758 & IV  & &&  2111$-$4335  & 1.708 & V   & \nl
1009$+$0222  & 1.349 & III & &&  2113$-$4345  & 2.053 & v   & 1,2 \nl
1016$-$0248  & 0.717 & III & &&  2116$-$4439  & 1.480 & V   & \nl
1029$-$0125  & 2.029 & III & &&  2140$-$4552  & 1.688 & V   & \nl
1133$+$0214  & 1.468 & III & &&  2201$-$1834  & 1.814 & V   & \nl
1138$-$0126  & 1.266 & III & &&  2208$-$1720  & 1.210 & V   & \nl
1203$+$1530  & 1.628 & III & &&  2211$-$1915  & 1.952 & V   & \nl
1205$+$1436  & 1.643 & I   & &&  2212$-$1759  & 2.217 & V   & \nl
1208$+$1535  & 1.961 & I   & &&  2241$+$0016  & 1.394 & II  & \nl
1212$+$1445  & 1.627 & I   & &&  2244$-$0105  & 2.030 & ii  & 1,2,7 \nl
1214$+$1753  & 0.679 & I   & &&  2350$-$0045A & 1.617 & II  & \nl
1216$+$1103  & 1.620 & I   & &&  2351$+$0120  & 2.068 &     & 1,2 \nl
1219$+$1244  & 1.309 & III & &&  2355$-$0209  & 1.802 & II  & \nl
1224$+$1349  & 1.838 & I   & &&  2358$+$0216  & 1.872 & IV  & \nl
1228$+$1216  & 1.408 & I   & &&  
\tablenotetext{a}{Original LBQS paper that identified BAL (upper case) 
or associated narrow absorption (lower case).}
\tablecomments{1) NAL on Lyman $\alpha$; 2) NAL on \ion{C}{4}; 
3) BAL near Lyman $\alpha$; 4) BAL near \ion{C}{4}; 
5) NAL on \ion{Mg}{2}; 6) BAL near \ion{Mg}{2}; 7) NAL on \ion{N}{5}.}
\tablerefs{(I,i) Foltz et al. 1987; (II,ii) Foltz et al. 1989; 
(III,iii) Hewett et al. 1991; (IV,iv) Chaffee et al. 1991; (V,v) 
Morris et al. 1991}
\enddata
\end{deluxetable}



\begin{deluxetable}{ccll}
\small
\tablewidth{330pt}
\tablenum{2}
\tablecaption{Continuum and iron fitting windows}
\tablehead{
\multicolumn{2}{c}{Rest frame wavelength range (\AA)} & 
\multicolumn{2}{c}{Emission lines nearby} \\
\colhead{Continuum} & \colhead{Iron} & \colhead{Blueward} & \colhead{Redward}
}
\startdata
1140 -- 1150\tablenotemark{a} &  & \ion{O}{6} $\lambda$1035 & Lyman $\alpha$ $\lambda$1215   \nl
1275 -- 1280   &   & \ion{N}{5} $\lambda$1243     & \ion{O}{1} $\lambda$1305   \nl
1320 -- 1330   &   & \ion{O}{1} $\lambda$1305     & \ion{Si}{4} + \ion{O}{4} $\lambda$1400\nl
1455 -- 1470   &   & \ion{Si}{4} + \ion{O}{4} $\lambda$1400 & \ion{C}{4} $\lambda$1549    \nl
1690 -- 1700   &   & \ion{He}{2} $\lambda$1640    & \ion{Al}{3} $\lambda$1859     \nl
2160 -- 2180   & 2020 -- 2120   &    \nl
2225 -- 2250   & 2250 -- 2650   & \raisebox{1.5ex}[0pt]{\ion{C}{3}] $\lambda$1909} & \raisebox{1.5ex}[0pt]{\ion{Mg}{2} $\lambda$2800}  \nl
3010 -- 3040\tablenotemark{b} & 2900 -- 3000   & \nl
3240 -- 3270   &   & \raisebox{1.5ex}[0pt]{\ion{Mg}{2} $\lambda$2800} & \raisebox{1.5ex}[0pt]{[\ion{Ne}{5}] $\lambda$3426}   \nl
3790 -- 3810   &   & [\ion{O}{2}] $\lambda$3728   & [\ion{Ne}{3}] $\lambda$3869 \nl
4210 -- 4230   &   & H$\delta$ $\lambda$4102    & H$\gamma$ $\lambda$4340  \nl
  & 4400 -- 4750\tablenotemark{c} & [\ion{O}{3}] $\lambda$4363 & H$\beta$ $\lambda$4861    \nl
5080 -- 5100   & 5150 -- 5500   & \nl
5600 -- 5630   &   & \raisebox{1.5ex}[0pt]{[\ion{O}{3}] $\lambda$5007}  & \raisebox{1.5ex}[0pt]{\ion{He}{1} $\lambda$5876}  \nl
5970 -- 6000   &   & \ion{He}{1} $\lambda$5876     & [\ion{N}{2}] $\lambda$6549 
\tablenotetext{a}{This window lies on the blue side of the Lyman $\alpha$ 
emission line and is only used where no other continuum window is available.}
\tablenotetext{b}{May have some iron emission contamination.}
\tablenotetext{c}{\ion{He}{2} $\lambda$4686 lies in this window.}
\enddata
\end{deluxetable}



\begin{deluxetable}{lcccl}
\small
\tablewidth{470pt}
\tablenum{3}
\tablecaption{Inventory of emission lines}
\tablehead{
              & $\lambda_{c}$ & FWHM$^{a}$    & Window$^{b}$  \nl
Emission line & (\AA)         & (km s$^{-1}$) & (\AA)  & Notes 
}
\startdata
Lyman $\beta$ $\lambda$1025.7 + \ion{O}{6} $\lambda$1035 & 1030.0 & 5000 & 1010 -- 1060 & See Appendix. \nl
Lyman $\alpha$ $\lambda$1215.7                           & 1215.7 & 7000s & 1170 -- 1350 & \nl
                                                         &        & 2500n, 8000b & \ldots      & \nl
\ion{N}{5} $\lambda$1241.5                               & 1241.5 & 6000  & \ldots       & \nl
\ion{O}{1} $\lambda$1305                                 & 1305.0 & 2500  & \ldots       & \nl
\ion{Si}{4} + \ion{O}{4} $\lambda$1400                   & 1400.0 & 5800  & 1350 -- 1450 & \nl
\ion{C}{4} $\lambda$1549                                & 1549.0 & 6500s & 1350 -- 1720 & \nl
                                                         &        & 3000n, 11000b & \ldots       & \nl
\ion{He}{2} $\lambda$1640                                & 1640.0 & 10000 & \ldots       & See Appendix. \nl
\ion{Al}{3} $\lambda$1859                                & 1859.0 & 3500  & 1820 -- 1970 & \nl
\ion{C}{3}] $\lambda$1909                                & 1909.0 & 5500  & \ldots       & \nl
\ion{Mg}{2} $\lambda$2800                                & 2800.0 & 4000s & 2700 -- 2900 & Iron emission may be strong in \nl
                                                         &        & 3000n, 8000b & \ldots & \quad this window.  \nl
[\ion{Ne}{5}] $\lambda$3426                              & 3426.0 & 1000  & 3390 -- 3460 & See Appendix. \nl
[\ion{O}{2}] $\lambda$3728                               & 3728.0 &  600  & 3700 -- 3760 & See Appendix. \nl
[\ion{Ne}{3}] $\lambda$3869                              & 3869.0 &  900  & 3810 -- 3930 & See Appendix. \nl
H$\delta$ $\lambda$4101.7                                & 4101.7 &  450  & 4000 -- 4200 & \nl
H$\gamma$ $\lambda$4340.5 + [\ion{O}{3}] $\lambda$4363   & 4352.0 & 3500  & 4240 -- 4440 & \nl
\ion{He}{2} $\lambda$4686.5                              & 4686.5 & 1200  & 4580 -- 4790 & See Appendix. \nl
H$\beta$ $\lambda$4861.3                                 & 4861.3 & 4000s & 4750 -- 5100 & Iron emission may be strong in\nl
                                                         &        & 1000n, 5500b & \ldots & \quad this window. \nl
[\ion{O}{3}] $\lambda$4959                    & 4959.0 &  700  & \ldots & Relative strength and separation of \nl
[\ion{O}{3}] $\lambda$5007                    & 5007.0 &  600  & \ldots & \quad [\ion{O}{3}] lines is fixed in 1st iteration. \nl
\ion{He}{1} $\lambda$5875.6                              & 5875.6 & 2000  & 5825 -- 5900 & 
\tablenotetext{a}{This is the FWHM used for estimating the 
upper limits of $W_{\lambda}$ in weak emission lines. More than one width is 
given for emission lines that can have two Gaussian components; 
s = single component, n = narrow component, b = broad component.}
\tablenotetext{b}{The wavelength range over which the emission lines 
are modeled.}
\enddata
\end{deluxetable}



\begin{deluxetable}{lccc}
\small
\tablewidth{260pt}
\tablenum{4}
\tablecaption{Total number of emission lines modeled}
\tablehead{
              & Fixed FWHM         & $W_{\lambda}$ & TOTAL \nl
Emission Line & and position\tablenotemark{a} & upper limits  & out of 993 
}
\startdata
UV iron                  & \ldots & 294 &  953 \nl
Optical iron             & \ldots & 109 &  247 \nl
Lyman $\beta$ + \ion{O}{6} &   10 &  27 &  130 \nl
{Lyman $\alpha$ \hfill single}    &    2 &   0 &  164 \nl
{\hfill narrow}    &    0 &   0 &   96 \nl
{\hfill broad}     &    0 &   0 &   96 \nl
\ion{N}{5}               &   24 &   8 &  260 \nl
\ion{O}{1}               &   20 & 121 &  260 \nl
\ion{Si}{4} + \ion{O}{4} &   12 &  19 &  414 \nl
{\ion{C}{4} \hfill single}        &    3 &   2 &  408 \nl
{\hfill narrow}        &    0 &   0 &   80 \nl
{\hfill broad}         &    0 &   0 &   80 \nl
\ion{He}{2} $\lambda$1640 &   66 &  69 &  488 \nl
\ion{Al}{3}              &   63 & 181 &  667 \nl
\ion{C}{3}]              &   20 &  26 &  667 \nl
{\ion{Mg}{2} \hfill single}       &    3 &  42 &  559 \nl
{\hfill narrow}       &    0 &   0 &  118 \nl
{\hfill broad}        &    0 &   0 &  118 \nl
[\ion{Ne}{5}]              &   10 & 365 &  488 \nl
[\ion{O}{2}]               &   27 & 272 &  393 \nl
[\ion{Ne}{3}]              &   12 & 259 &  363 \nl
H $\delta$               &    6 & 222 &  309 \nl
H $\gamma$ + [\ion{O}{3}]&    4 &  87 &  251 \nl
\ion{He}{2} $\lambda$4686 &    5 & 155 &  187 \nl
{H $\beta$ \hfill single}         &    4 &  18 &  141 \nl
{\hfill narrow}         &    0 &   0 &    7 \nl
{\hfill broad}          &    0 &   0 &    7 \nl
[\ion{O}{3}] $\lambda$4959 &    6 &  73 &  148 \nl
[\ion{O}{3}] $\lambda$5007 &    8 &  54 &  148 \nl
\ion{He}{1}              &    0 &  30 &   33 \nl
\noalign{\vskip 0.2cm}
{\hfill TOTALS}          &  305 & 2437 & 8288 
\tablenotetext{a}{Does not include $W_{\lambda}$ upper limits.}
\enddata
\end{deluxetable}



\begin{deluxetable}{lrrrc}
\small
\tablewidth{320pt}
\tablenum{5}
\tablecaption{Representative emission line measurements}
\tablehead{
\noalign{\vskip 0.1cm}
Designation \qquad (z) \nl
\noalign{\vskip 0.1cm}
& {\hfill FWHM \hfill} & 
{\hfill $\Delta \lambda_{p}$ \hfill} & 
{\hfill $W_{\lambda}$ \hfill} & Absorption \nl 
\noalign{\vskip 0.05cm}
\quad Emission Line & {\hfill (km s$^{-1}$) \hfill} & 
{\hfill (km s$^{-1}$) \hfill} & 
{\hfill (\AA) \hfill} & Lines \nl
\noalign{\vskip 0.05cm}
{\hfill (1) \hfill} & {\hfill (2) \hfill} & 
{\hfill (3) \hfill} & {\hfill (4) \hfill} & {\hfill (5) \hfill} 
}
\startdata
\noalign{\vskip 0.05cm}
0025+0009 \quad z = 0.205  \nl
\noalign{\vskip 0.05cm}
\quad UV iron                        & $  900 ^{+9000}_{-250}$ &  {\hfill \ldots \hfill} & $ 70 ^{+65}_{-65}$ & \ldots \nl
\noalign{\vskip 0.05cm}
\quad Optical iron                   & $ 1200 ^{+8750}_{-250}$ &  {\hfill \ldots \hfill} & $ 74 ^{+15}_{-15}$ & \ldots \nl
\noalign{\vskip 0.05cm}
\quad \ion{Mg}{2} \hfill single      & $ 2800 ^{+1000}_{-850}$ & $-500 ^{+440}_{-440}$ & $ 40 ^{+35}_{-30}$ & \ldots \nl
\noalign{\vskip 0.05cm}
\quad [\ion{Ne}{5}]                    & $ 1200 ^{+1100}_{-750}$ & $-100 ^{+380}_{-380}$ & $  5.5 ^{+10.0}_{-4.3}$ & \ldots \nl
\noalign{\vskip 0.05cm}
\quad [\ion{O}{2}]                     & $  210 ^{+440}_{-200}$ & $100 ^{+140}_{-140}$ & $  3.6 ^{+6.1}_{-3.1}$ & \ldots \nl
\noalign{\vskip 0.05cm}
\quad [\ion{Ne}{3}]                    & $ 1000 ^{+3000}_{-700}$ & $ 0 ^{+2200}_{-2200}$ & $  5.0 ^{+30.0}_{-4.4}$ & \ldots \nl
\noalign{\vskip 0.05cm}
\quad H $\delta$                     & $ 5500 ^{+5800}_{-1100}$ & $0 ^{+1100}_{-1000}$ & $ 35 ^{+55}_{-14}$ & \ldots \nl
\noalign{\vskip 0.05cm}
\quad H $\gamma$ + [\ion{O}{3}]      & $ 2800 ^{+850}_{-700}$ & $ -200 ^{+380}_{-380}$ & $ 34 ^{+20}_{-14}$ & 1 \nl
\noalign{\vskip 0.05cm}
\quad \ion{He}{2} $\lambda$4686      & {\hfill $1200$  \hfill} & {\hfill \ldots \hfill} & {$\le 15$ \hfill} & \ldots \nl
\noalign{\vskip 0.05cm}
\quad H $\beta$ \hfill single        & $ 3000 ^{+550}_{-500}$ & $  300 ^{+240}_{-240}$ & $105 ^{+35}_{-30}$ & \ldots \nl
\noalign{\vskip 0.05cm}
\quad [\ion{O}{3}] $\lambda$4959       & $  600 ^{+1100}_{-280}$ & $    0 ^{+550}_{-550}$ & $ 15.0 ^{+40.0}_{-8.5}$ & \ldots \nl
\noalign{\vskip 0.05cm}
\quad [\ion{O}{3}] $\lambda$5007       & $  450 ^{+120}_{-120}$ & $  100 ^{+ 50}_{-50}$ & $ 33 ^{+13}_{-11}$ & \ldots \nl
\noalign{\vskip 0.05cm}
\quad \ion{He}{1}                    & $  600 ^{+650}_{-460}$ & $ -300 ^{+240}_{-240}$ & $  8.0 ^{+25.0}_{-7.9}$ & \ldots \nl
\noalign{\vskip 0.2cm}
2244+0020 \quad z = 0.973 \nl
\noalign{\vskip 0.05cm}
\quad UV iron                        & $ 4250 ^{+5750}_{-3250}$ &  {\hfill \ldots \hfill} & $ 50.6 ^{+1.7}_{-1.7}$ & \ldots \nl
\noalign{\vskip 0.05cm}
\quad \ion{Al}{3}                    & $ 6400 ^{+600}_{-500}$ & $ 900 ^{+280}_{-280}$ & $ 12.8 ^{+4.7}_{-4.4}$ & \ldots \nl
\noalign{\vskip 0.05cm}
\quad \ion{C}{3}]                     & $ 5550 ^{+180}_{-160}$ & $ -400 ^{+90}_{-90}$ & $ 27.9 ^{+7.3}_{-7.2}$ & \ldots \nl
\noalign{\vskip 0.05cm}
\quad \ion{Mg}{2} \hfill narrow      & $ 4000 ^{+280}_{-260}$ & $ -250 ^{+120}_{-140}$ & $ 35.6 ^{+4.7}_{-4.0}$ & \ldots \nl
\noalign{\vskip 0.05cm}
                  \hfill broad       & $12500 ^{+2100}_{-1500}$ & $  200 ^{+900}_{-900}$ & $ 27.4 ^{+8.5}_{-6.3}$ & \ldots \nl
\noalign{\vskip 0.05cm}
\quad [\ion{Ne}{5}]                    & {\hfill $1000$ \hfill}  & {\hfill \ldots \hfill} & {$\le 2.8$ \hfill} & \ldots \nl
\noalign{\vskip 0.2cm}
2354$-$0134 \quad z = 2.211  \nl
\noalign{\vskip 0.05cm}
\quad UV iron                        & $ 2000 ^{+8000}_{-1000}$ &  {\hfill \ldots \hfill} & $ 35 ^{+25}_{-25}$ & \ldots \nl
\noalign{\vskip 0.05cm}
\quad Lyman $\beta$ + \ion{O}{6}     & $ 5400 ^{+850}_{-800}$ & $ 1300 ^{+420}_{-420}$ & $ 24.9 ^{+7.8}_{-6.5}$ & \ldots \nl
\noalign{\vskip 0.05cm}
\quad Lyman $\alpha$ \hfill narrow   & $ 3500 ^{+200}_{-200}$ & $  150 ^{+ 90}_{-140}$ & $ 33 ^{+11}_{-10}$ & \ldots \nl
\noalign{\vskip 0.05cm}
                     \hfill broad    & $ 8100 ^{+440}_{-340}$ & $  500 ^{+180}_{-240}$ & $ 57 ^{+17}_{-17}$ & 1 \nl
\noalign{\vskip 0.05cm}
\quad \ion{N}{5}                     & $ 5400 ^{+400}_{-380}$ & $    0 ^{+160}_{-220}$ & $ 34 ^{+11}_{-11}$ & \ldots \nl
\noalign{\vskip 0.05cm}
\quad \ion{O}{1}                     & $  600 ^{+420}_{-440}$ & $  100 ^{+160}_{-160}$ & $  1.3 ^{+1.5}_{-0.8}$ & 1 \nl
\noalign{\vskip 0.05cm}
\quad \ion{Si}{4} + \ion{O}{4}       & $ 5000 ^{+700}_{-650}$ & $    0 ^{+380}_{-380}$ & $ 13.4 ^{+3.8}_{-3.2}$ & \ldots \nl
\noalign{\vskip 0.05cm}
\quad \ion{C}{4} \hfill single       & $ 6100 ^{+360}_{-360}$ & $   50 ^{+180}_{-180}$ & $ 52.7 ^{+6.0}_{-5.7}$ & \ldots \nl
\noalign{\vskip 0.05cm}
\quad \ion{He}{2} $\lambda$1640      & $17000 ^{+3800}_{-3100}$ & $-1200 ^{+650}_{-650}$ & $ 26.0 ^{+12.0}_{-8.6}$ & \ldots \nl
\noalign{\vskip 0.05cm}
\quad \ion{Al}{3}                    & {\hfill $3500$ \hfill} & {\hfill \ldots \hfill} & {$\le 7.0$ \hfill} & \ldots \nl
\noalign{\vskip 0.05cm}
\quad \ion{C}{3}]                     & $ 6500 ^{+1900}_{-1500}$ & $ -400 ^{+900}_{-600}$ & $ 32 ^{+17}_{-12}$ & \ldots
\enddata
\tablecomments{This is a digested form of the table available from the 
electronic journal. 
Only emission lines measured in three example LBQS spectra are presented 
here, see \S4 for more details.}
\end{deluxetable}



\begin{deluxetable}{lccrcrrc}
\small
\tablewidth{380pt}
\tablenum{6}
\tablecaption{Representative continuum measurements}
\tablehead{
&& & \multicolumn{4}{c}{Power law continua} & 
Polynomial \nl
Designation & Redshift & N$_{\rm H}^{\rm Gal}$  & 
{\hfill $\Gamma_{1}$ \hfill} & 
$\lambda_{c}$ & {\hfill $A_{\lambda}$ \hfill} & {\hfill $\Gamma_{2}$ \hfill} 
& Order \nl
\noalign{\vskip 0.05cm}
{\hfill (1) \hfill} & {\hfill (2) \hfill} & 
{\hfill (3) \hfill} & {\hfill (4) \hfill} & 
{\hfill (5) \hfill} & {\hfill (6) \hfill} & 
{\hfill (7) \hfill} & {\hfill (8) \hfill} 
}
\startdata
0025+0009 & 0.205 & 3.010 & 3.99 $^{+1.37}_{-1.51}$ & 4220 & 
1.27 $^{+0.16}_{-0.15}$ & 1.08 $^{+0.48}_{-0.48}$ & 2  \nl
\noalign{\vskip 0.05cm}
2244+0020 & 0.973 & 5.318 & 2.50 $^{+0.01}_{-0.01}$ & 2170 & 
15.4 $^{+0.9}_{-0.1}$ & {\hfill \ldots \hfill} & 2  \nl
\noalign{\vskip 0.05cm}
2354$-$0134 & 2.211 & 3.368 & 2.45 $^{+0.47}_{-0.24}$ & 1463 & 
1.19 $^{+0.04}_{-0.08}$ & {\hfill \ldots \hfill} & 1  \nl
\enddata
\tablecomments{This is a digested form of the table available from 
the electronic journal. Only the continuum parameters from three  
example LBQS spectra are presented here, see \S4 for more details. The
units for the normalization of the power law continuum ($A_{\lambda}$)
presented here are $10^{-16}$ ergs cm$^{-2}$ s$^{-1}$ \AA$^{-1}$. In the
electronic version of this table the normalization factor is $10^{-14}$.}
\end{deluxetable}



\begin{deluxetable}{lrrrcrrrrcrrr}
\small
\tablewidth{520pt}
\tablenum{7}
\tablecaption{Emission line parameter distributions for the LBQS sample}
\tablehead{
& \multicolumn{8}{c}{$W_{\lambda}$} &&
\multicolumn{3}{c}{FWHM} \nl
\cline{2-9} \cline{11-13} \nl
& \multicolumn{3}{c}{Detected} & \quad & 
\multicolumn{4}{c}{Kaplan-Meier} & \quad & 
\multicolumn{3}{c}{Detected} \nl
Emission Line & Num & Mean & {\hfill SD \hfill}  && Num & Limits & 
{\hfill Mean \hfill} & Median && Num & 
{\hfill Mean \hfill} & Median \nl
{\hfill (1) \hfill} & {\hfill (2) \hfill} & 
{\hfill (3) \hfill} & {\hfill (4) \hfill} && 
{\hfill (5) \hfill} & {\hfill (6) \hfill} & 
{\hfill (7) \hfill} & {\hfill (8) \hfill} && 
{\hfill (9) \hfill} & {\hfill (10) \hfill} & 
{\hfill (11) \hfill}  
}
\startdata
UV iron                           &  659 &   40.2 &   23.0 &&  953 &  294 &   29.9 $\pm$    0.8 &   27.3 &&  659 &  4610 $\pm$   130 &  3470 \nl
Optical iron                      &  138 &   39.0 &   21.5 &&  247 &  109 &   23.8 $\pm$    1.6 &   21.3 &&  138 &  4630 $\pm$   310 &  2600 \nl
Lyman $\beta$ + \ion{O}{6}        &  103 &   11.7 &   10.1 &&  130 &   27 &    9.5 $\pm$    0.9 &    7.3 &&   93 &  4870 $\pm$   280 &  4410 \nl
{Lyman $\alpha$ \hfill single}\tablenotemark{a}    &  164 &   56.4 &   28.8 &&  164 &    0 &   56.4 $\pm$    2.2 &   53.0 &&  162 &  7820 $\pm$   230 &  7650 \nl
{\hfill narrow}                   &   96 &   28.1 &   21.3 &&   96 & \ldots &   28.1 $\pm$    2.2 &   22.0 &&   96 &  2650 $\pm$   120 &  2470 \nl
{\hfill broad}                    &   96 &   68.3 &   31.6 &&   96 & \ldots &   68.3 $\pm$    3.2 &   64.0 &&   96 & 10250 $\pm$   420 &  9330 \nl
{\hfill sum}\tablenotemark{b}                      &  260 &   71.2 &   40.4 &&  260 &    0 &   71.2 $\pm$    2.5 &   61.0 &&  \ldots & {\hfill \ldots \hfill} & \ldots \nl
\ion{N}{5}                        &  252 &   18.8 &   10.7 &&  260 &    8 &   18.3 $\pm$    0.7 &   16.5 &&  228 &  5580 $\pm$   170 &  5400 \nl
\ion{O}{1}                        &  139 &    3.1 &    2.5 &&  260 &  121 &    2.0 $\pm$    0.1 &    1.4 &&  119 &  2990 $\pm$   190 &  2460 \nl
\ion{Si}{4} + \ion{O}{4}          &  395 &   13.0 &    8.8 &&  414 &   19 &   12.6 $\pm$    0.4 &   10.9 &&  383 &  6780 $\pm$   160 &  5940 \nl
{\ion{C}{4} \hfill single}\tablenotemark{a}        &  406 &   38.1 &   19.3 &&  408 &    2 &   38.0 $\pm$    1.0 &   34.5 &&  403 &  7720 $\pm$   150 &  7300 \nl
{\hfill narrow}                   &   80 &   17.8 &    9.3 &&   80 & \ldots &   17.8 $\pm$    1.0 &   15.5 &&   80 &  2860 $\pm$   110 &  2750 \nl
{\hfill broad}                    &   80 &   44.8 &   23.1 &&   80 & \ldots &   44.8 $\pm$    2.6 &   37.4 &&   80 & 10960 $\pm$   360 & 10530 \nl
{\hfill sum}\tablenotemark{b}                      &  486 &   42.1 &   23.1 &&  488 &    2 &   42.0 $\pm$    1.0 &   37.0 &&  \ldots & {\hfill \ldots \hfill} & \ldots \nl
\ion{He}{2} $\lambda$1640         &  419 &   18.7 &   13.5 &&  488 &   69 &   16.4 $\pm$    0.6 &   12.9 &&  353 & 14670 $\pm$   390 & 13590 \nl
\ion{Al}{3}                       &  486 &    8.7 &    6.0 &&  667 &  181 &    6.8 $\pm$    0.2 &    5.2 &&  423 &  5740 $\pm$   130 &  5620 \nl
\ion{C}{3}]                       &  641 &   28.4 &   14.9 &&  667 &   26 &   27.6 $\pm$    0.6 &   24.8 &&  621 &  7820 $\pm$   170 &  6670 \nl
{\ion{Mg}{2} \hfill single}\tablenotemark{a}       &  517 &   39.1 &   21.4 &&  559 &   42 &   37.1 $\pm$    0.9 &   33.8 &&  514 &  5160 $\pm$   120 &  4440 \nl
{\hfill narrow}                   &  118 &   28.1 &   12.9 &&  118 & \ldots &   28.1 $\pm$    1.2 &   26.0 &&  118 &  3510 $\pm$   110 &  3370 \nl
{\hfill broad}                    &  118 &   36.6 &   21.5 &&  118 & \ldots &   36.6 $\pm$    2.0 &   31.6 &&  118 &  8660 $\pm$   300 &  8880 \nl
{\hfill sum}\tablenotemark{b}                      &  635 &   43.9 &   25.1 &&  677 &   42 &   42.0 $\pm$    1.0 &   36.8 &&  \ldots & {\hfill \ldots \hfill} & \ldots \nl
[\ion{Ne}{5}]                     &  123 &    5.8 &    5.8 &&  488 &  365 &    2.1 $\pm$    0.2 &    0.4 &&  113 &  1570 $\pm$ \hskip 0.14cm    90 &  1420 \nl
[\ion{O}{2}]                      &  121 &    7.8 &   14.6 &&  393 &  272 &    3.2 $\pm$    0.4 &    1.0 &&   94 &   900 $\pm$ \hskip 0.14cm    70 &   610 \nl
[\ion{Ne}{3}]                     &  104 &    7.5 &    5.0 &&  363 &  259 &    3.4 $\pm$    0.2 &    1.8 &&   92 &  1770 $\pm$   130 &  1380 \nl
H$\delta$                         &   87 &   12.3 &    8.6 &&  309 &  222 &    3.8 $\pm$    0.4 &    0.1 &&   81 &  2860 $\pm$   220 &  2300 \nl
H$\gamma$ + [\ion{O}{3}]          &  164 &   18.3 &   11.1 &&  251 &   87 &   14.6 $\pm$    0.7 &   11.8 &&  160 &  2920 $\pm$   140 &  2500 \nl
\ion{He}{2} $\lambda$4686         &   32 &   21.0 &   20.1 &&  187 &  155 &    4.2 $\pm$    0.9 &    0.1 &&   27 &  3690 $\pm$   550 &  2350 \nl
{H$\beta$ \hfill single}\tablenotemark{a}          &  123 &   62.4 &   36.0 &&  141 &   18 &   57.4 $\pm$    3.1 &   49.8 &&  119 &  4370 $\pm$   250 &  3760 \nl
{\hfill narrow}                   &    7 &   34.3 &   13.6 &&    7 & \ldots &   34.3 $\pm$    4.8 &   31.4 &&    7 &  1160 $\pm$   270 &   900 \nl
{\hfill broad}                    &    7 &  101.6 &   58.9 &&    7 & \ldots &  101.6 $\pm$    0.6 &   77.5 &&    7 &  6560 $\pm$   850 &  5300 \nl
{\hfill sum}\tablenotemark{b}                      &  130 &   66.4 &   41.0 &&  148 &   18 &   61.2 $\pm$    3.4 &   53.8 &&  \ldots & {\hfill \ldots \hfill} & \ldots \nl
[\ion{O}{3}] $\lambda$4959        &   75 &   13.8 &   14.0 &&  146 &   71 &    8.5 $\pm$    1.0 &    5.5 &&   69 &   940 $\pm$ \hskip 0.14cm    80 &   770 \nl
[\ion{O}{3}] $\lambda$5007        &   94 &   30.4 &   39.0 &&  146 &   52 &   21.6 $\pm$    2.8 &   13.7 &&   86 &   820 $\pm$ \hskip 0.14cm    70 &   590 \nl
\ion{He}{1}                       &    3 &   10.7 &    3.8 &&   33 &   30 &    4.7 $\pm$    1.1 &    1.7 &&    3 &   820 $\pm$   120 &  750
\enddata
\tablenotetext{a}{The distribution for single Gaussian component models 
are tabulated separately from narrow and broad components.}
\tablenotetext{b}{The distribution of sum of the broad and narrow
component \ew~ included with the single component measurements (see \S5).}
\tablecomments{(1) Emission line or line blend. (2) Number of detected
emission lines, (3) Mean \ew~ of detected emission lines, (4) Standard
deviation (SD) of \ew~ measurements for detected emission lines. (5) Total
number of measured emission lines, (6) The number of upper limits
estimates of \ew, (7)--(8) The KM reconstructed mean and median of the
\ew~ distributions (see \S5). (9)--(11) The number, mean and median of
the distribution of FWHM of the Gaussian components used to model each
emission feature. All \ew~ are rest frame measurements in \AA~ and the
FWHM are in km s$^{-1}$.}
\end{deluxetable}



\begin{deluxetable}{lcccrrccrr}
\small
\tablewidth{430pt}
\tablenum{8}
\tablecaption{Emission line parameter distributions for published comparison samples}
\tablehead{
&& \multicolumn{4}{c}{$W_{\lambda}$} && 
\multicolumn{3}{c}{FWHM} \nl
Emission Line & Ref & Num & Limits &  {\hfill Mean \hfill} & Median & \qquad & Num &  {\hfill Mean \hfill} & Median \nl
{\hfill (1) \hfill} & {\hfill (2) \hfill} & 
{\hfill (3) \hfill} & {\hfill (4) \hfill} & 
{\hfill (5) \hfill} & {\hfill (6) \hfill} && 
{\hfill (7) \hfill} & {\hfill (8) \hfill} & 
{\hfill (9) \hfill} 
}
\startdata
UV iron           & 2 &   38 &    8 &   28.1 $\pm$    4.8\tablenotemark{a} &   21.5 && \ldots & {\hfill \ldots \hfill} &  \ldots  \nl
Optical iron      & 1 &   87 &    7 &   47.5 $\pm$    2.8 &   43.8 && \ldots & {\hfill \ldots \hfill} &  \ldots  \nl
                  & 2 &   45 &    5 &  178.5 $\pm$    6.2 &  110.9 && \ldots & {\hfill \ldots \hfill} &  \ldots  \nl
& 3Q & 14 & \ldots & 25.6 $\pm$ 4.9 & \ldots &&  \ldots & {\hfill \ldots \hfill} & \ldots \nl
& 3L & 17 & \ldots & 34.4 $\pm$ 4.2 & \ldots &&  \ldots & {\hfill \ldots \hfill} & \ldots \nl
& 4Q & 80 & \ldots & 52 $\pm$ 2.8\tablenotemark{b} & \ldots &&  \ldots & {\hfill \ldots \hfill} & \ldots \nl
& 4L & 45 & \ldots & 21.5 $\pm$ 2.5\tablenotemark{b} & \ldots &&  \ldots & {\hfill \ldots \hfill} & \ldots \nl
Lyman $\alpha$ & 3Q & 11 & \ldots & 72.2 $\pm$ 9.0 & \ldots && \ldots & {\hfill \ldots \hfill} & \ldots \nl
               & 3L & 6  & \ldots & 99.2 $\pm$ 9.0 & \ldots && \ldots & {\hfill \ldots \hfill} & \ldots \nl
{\ion{C}{4} \hfill single} & 3Q & 13 & \ldots & 22.6 $\pm$ 3.4 & \ldots && 15 & 4880 $\pm$ 460 & \ldots \nl
                            & 3L & 13 & \ldots & 20.0 $\pm$ 2.5 & \ldots && 15 & 6020 $\pm$ 670 & \ldots \nl
{\hfill narrow}   & 2 &   39 &    0 &   11.2 $\pm$    1.4 &    8.7 &&   39 &  1560 $\pm$ \hskip 0.14cm    80 &  1560 \nl
{\hfill broad}                    & 2 &   52 &    0 &  107.5 $\pm$    9.4 &   96.0 &&   51 &  6990 $\pm$   310 &  6610 \nl
\ion{He}{2} $\lambda$1640         & 2 &   44 &    3 &   25.3 $\pm$    2.7 &   23.4 &&   24 & 12940 $\pm$   600 & 12400 \nl
\ion{C}{3}] $\lambda$1909 & 3Q & 10 & \ldots & 17.9 $\pm$ 1.6 & \ldots && 12 & 6430 $\pm$ 730 & \ldots \nl
                          & 3L & 13 & \ldots & 17.9 $\pm$ 0.7 & \ldots && 14 & 8420 $\pm$ 620 & \ldots \nl
\ion{Mg}{2} $\lambda$2800 & 3Q & 7 & \ldots & 23.1 $\pm$ 2.3 & \ldots && \ldots & {\hfill \ldots \hfill} & \ldots \nl
                          & 3L & 6 & \ldots & 29.1 $\pm$ 4.1 & \ldots && \ldots & {\hfill \ldots \hfill} & \ldots \nl
\ion{He}{2} $\lambda$4686         & 1 &   70 &    0 &   11.2 $\pm$    1.1 &    7.5 && \ldots & {\hfill \ldots \hfill} &  \ldots  \nl
                                  & 2 &   32 &    5 &   16.8 $\pm$    2.4 &   13.5 &&   14 &  7480 $\pm$   770 &  7900 \nl
{H$\beta$ \hfill single}          & 1 &   87 &    0 &   95.5 $\pm$    4.0 &   92.5 &&   87 &  3790 $\pm$   220 &  3160 \nl
& 3Q & 14 & \ldots & 64.8 $\pm$ 6.9 & \ldots && 14 & 4430 $\pm$ 590 & \ldots \nl
& 3L & 17 & \ldots & 73.2 $\pm$ 5.1 & \ldots && 17 & 5100 $\pm$ 470 & \ldots \nl
& 4Q & 80 & \ldots & 99  $\pm$ 4.2\tablenotemark{b} & \ldots && \ldots & {\hfill \ldots \hfill} & \ldots \nl
& 4L & 45 & \ldots & 84 $\pm$ 5.2\tablenotemark{b} & \ldots && \ldots & {\hfill \ldots \hfill} & \ldots \nl
{\hfill narrow}     & 2 &   42 &    0 &    4.9 $\pm$    0.6 &    4.0 &&   38 &   630 $\pm$ \hskip 0.14cm    40 &   590 \nl
{\hfill broad}                  & 2 &   52 &    0 &  105.3 $\pm$    8.4 &  107.0 &&   51 &  5950 $\pm$   550 &  4780 \nl
& 3Q & \ldots & \ldots & {\hfill \ldots \hfill} & \ldots && 14 & 9870 $\pm$ 970 & \ldots \nl
& 3L & \ldots & \ldots & {\hfill \ldots \hfill} & \ldots && 17 & 11890 $\pm$ 600 & \ldots \nl
[\ion{O}{3}] $\lambda$5007 & 1 &   83 &  0 &   24.8 $\pm$    2.7 &   16.8 && \ldots & {\hfill \ldots \hfill} &  \ldots  \nl
                     & 2 &   49 &    0 &   58.8 $\pm$    6.8 &   23.4 &&   49 &   700 $\pm$ \hskip 0.14cm    40 &   610 \nl
& 3Q & 15 & \ldots & 20.6 $\pm$ 3.5 & \ldots && 14 & 1160 $\pm$ \hskip 0.14cm 90 & \ldots \nl
& 3L & 17 & \ldots & 10.2 $\pm$ 1.9 & \ldots && 15 & 1150 $\pm$ 130 & \ldots \nl
& 4Q & 80 & \ldots & 23.5 $\pm$ 2.9\tablenotemark{b} & \ldots && \ldots & {\hfill \ldots \hfill} & \ldots \nl
& 4L & 45 & \ldots & 23.0 $\pm$ 2.2\tablenotemark{b} & \ldots && \ldots & {\hfill \ldots \hfill} & \ldots
\enddata
\tablenotetext{a}{This measurement is not suitable for direct
comparison with the mean from the LBQS sample (Table 8) because it is
measured in a different region of the spectrum (see \S6).}
\tablenotetext{b}{Errors on mean calculated from $\sigma$/$\sqrt{N}$.}
\tablecomments{(1) Emission line or line blend. (2) Reference.  (3)
The total number of measured emission lines, (4) Number of upper
limits on \ew.  (5)--(6) The mean and median of the \ew~
distributions. Where upper limits are present the KM estimator was
used (see \S5). (7)--(9) The number, mean and median of the
distribution of published FWHM. All \ew~ are rest frame measurements
in \AA~ and the FWHM are in km s$^{-1}$.}  
\tablerefs{(1) Boroson \& Green 1992; (2) Marziani et al. 1996; (3)
McIntosh et al. 1999; (4) Sulentic et al. (2000); Q = Radio Quiet, L =
Radio Loud}
\end{deluxetable}


\clearpage


\begin{figure}[h]
\figurenum{1}
\plotfiddle{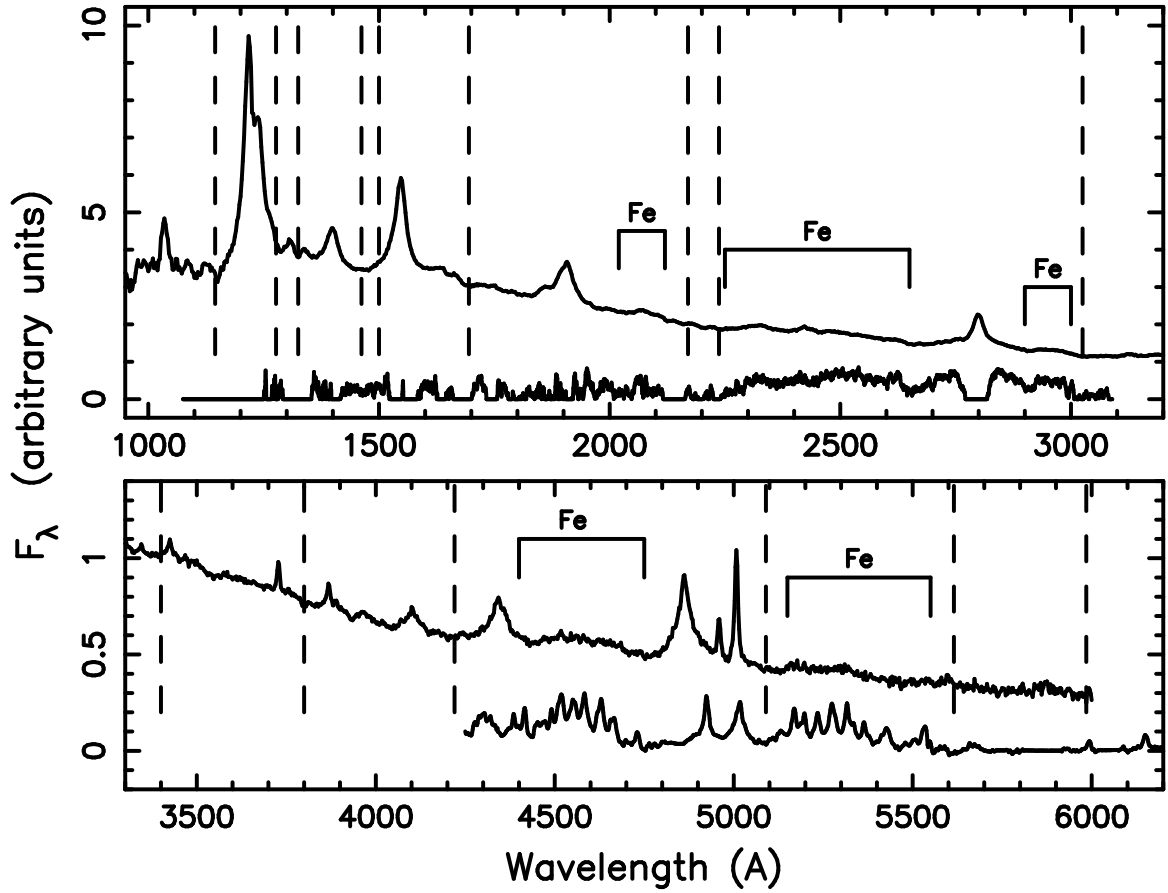}{12cm}{90}{60}{60}{240}{-20}
\caption{Continuum and iron template modeling windows. The composite
QSO spectrum from Francis et al. (1991) is shown above the iron
emission templates for the UV (upper panel) and optical (lower panel)
regions. The vertical dashed lines mark the positions of the narrow
windows used to constrain the continuum shape for each spectrum. 
The spectral windows used in the modeling of iron emission
are shown. See \S\S3.1, 3.2 and Table 2 for further details.}
\end{figure}

\begin{figure}[h]
\figurenum{2}
\plotfiddle{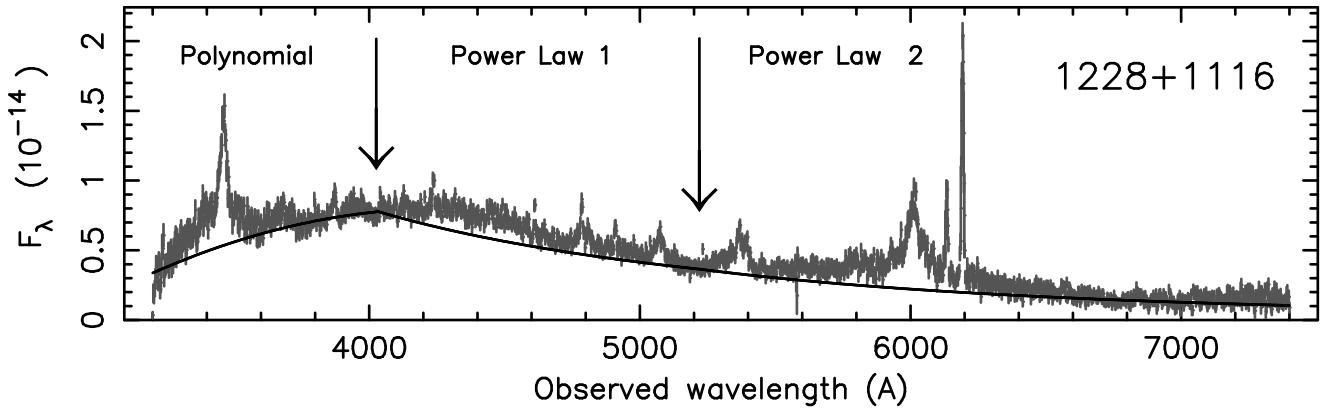}{6cm}{270}{70}{70}{-270}{400}
\caption{The continuum model for LBQS 1228+1116 (z=0.237). The bond
between the polynomial continuum and the first power law is at 4000\AA~ in the
{\em observed} frame and the inflection point between the two power law
continua is at {\em rest frame} 4220\AA~ (see \S3.1).}
\end{figure}

\clearpage
\eject

\begin{figure}[h]
\figurenum{3a}
\plotfiddle{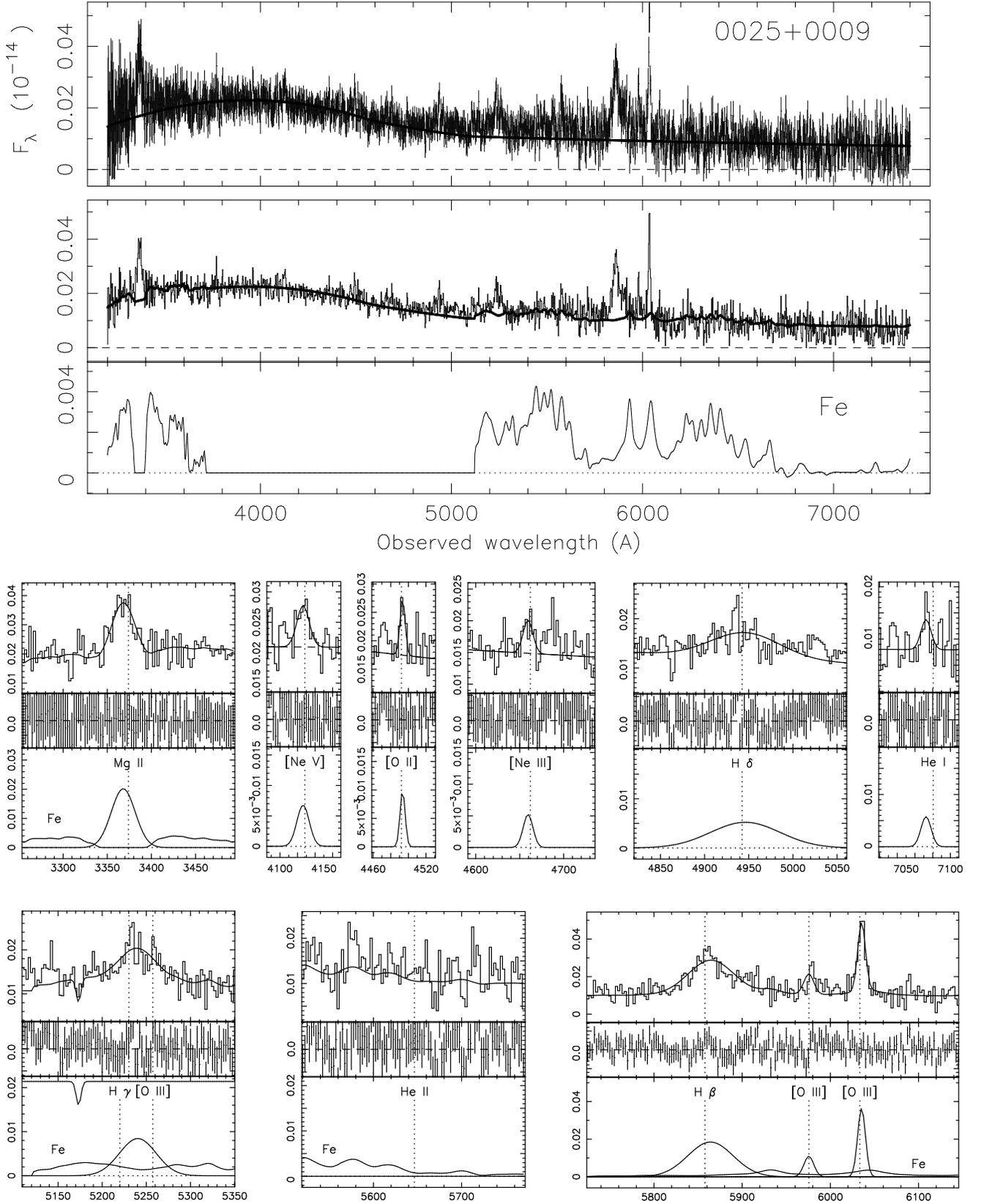}{21.5cm}{0}{90}{90}{-270}{-50}
\caption{The spectral models of three QSOs from the LBQS sample.
For each QSO, the top panel shows 3 sections.
The continuum model plotted over the
observed spectrum including error bars on each bin, the continuum
$+$ iron emission template plotted over the spectrum, and the iron
template profile alone. Note that for clarity the flux
scales are different in these sections. Smaller panels 
for each emission line component show the total best-fit 
model plotted over the relevant region of each spectrum, the residuals, and 
the individual Gaussian components (and the profile of the iron template 
emission). Flux units are 
$10^{-14}$ ergs cm$^{-2}$s$^{-1}$ \AA$^{-1}$, wavelength   
units are in \AA~ and are observed frame values. 
Broad and narrow components of the same emission line species are 
marked `b' or `n' respectively.
See \S4 for more details. (a) --- The spectral model of LBQS 0025$+$0009.}
\end{figure}
 
\clearpage
\eject

\begin{figure}[h]
\figurenum{3b}
\plotfiddle{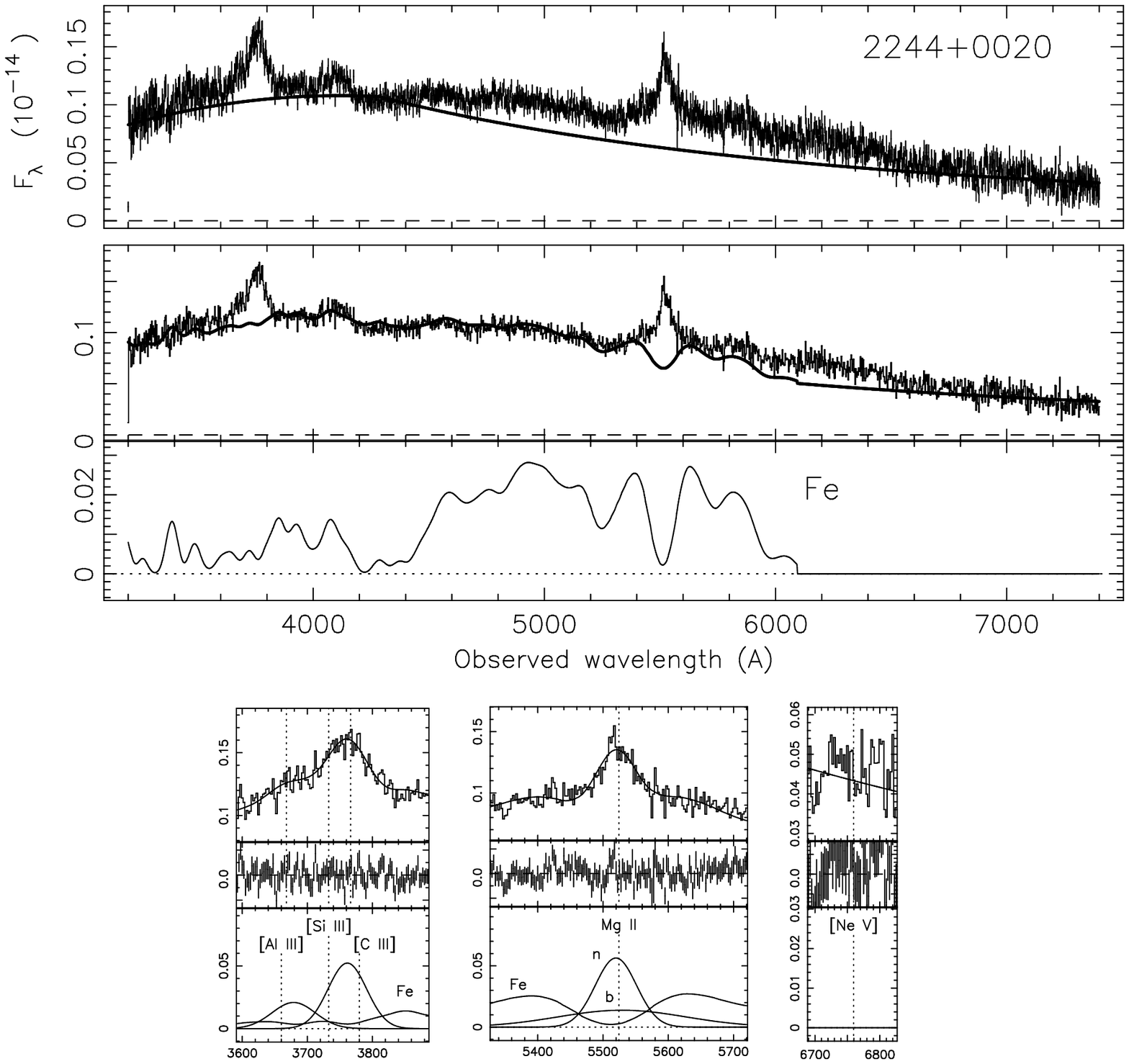}{19cm}{0}{90}{90}{-270}{-210}
\caption{The spectral model of LBQS 2244+0020 --- See Fig.3a and \S4.}
\end{figure}
 
\begin{figure}[h]
\figurenum{3c}
\plotfiddle{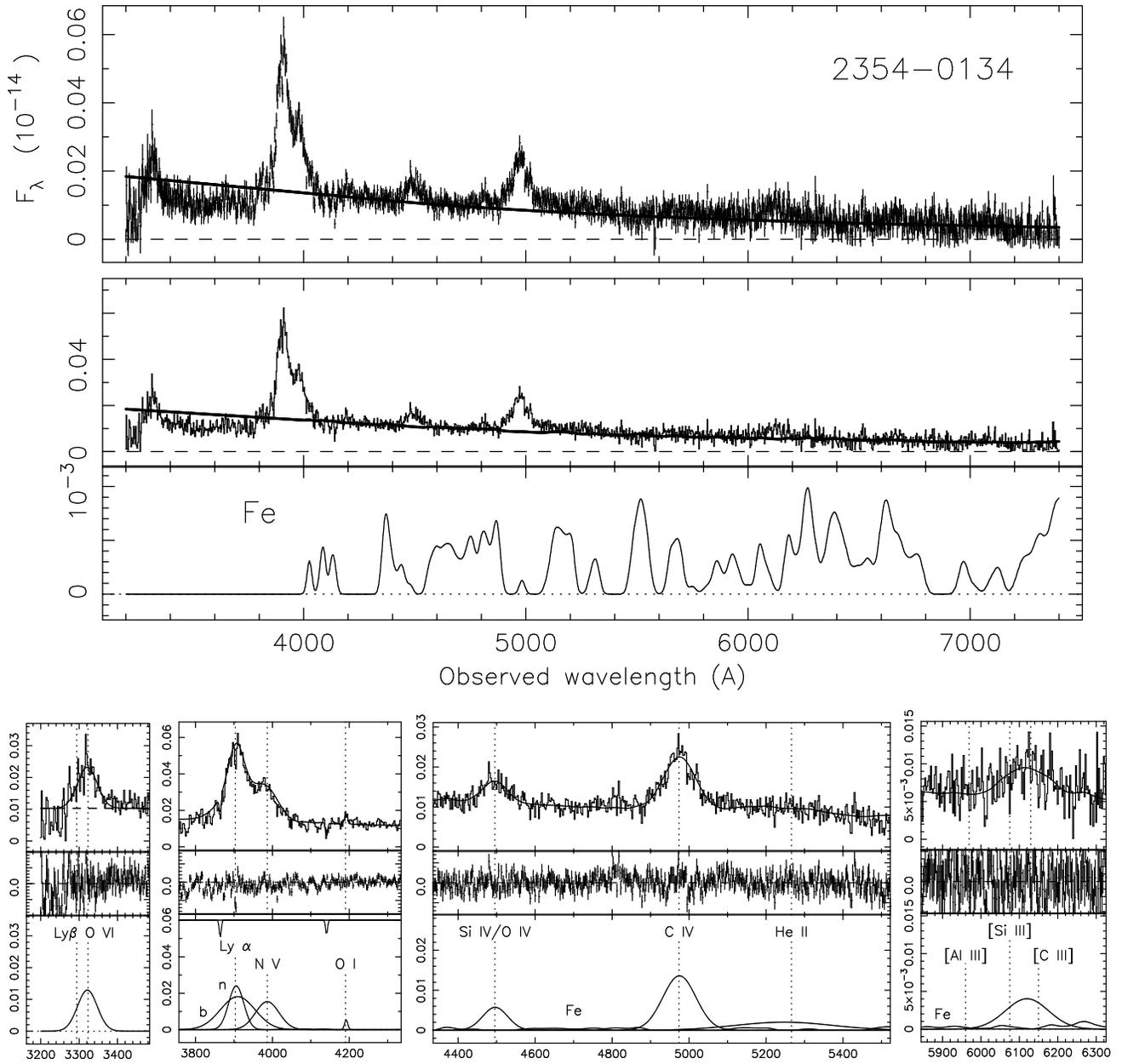}{19cm}{0}{90}{90}{-270}{-210}
\caption{The spectral model of LBQS 2354$-$0134 --- See Fig.3a and \S4.}
\end{figure}
 
\clearpage
\eject

\begin{figure}[h]
\figurenum{4}
\plotfiddle{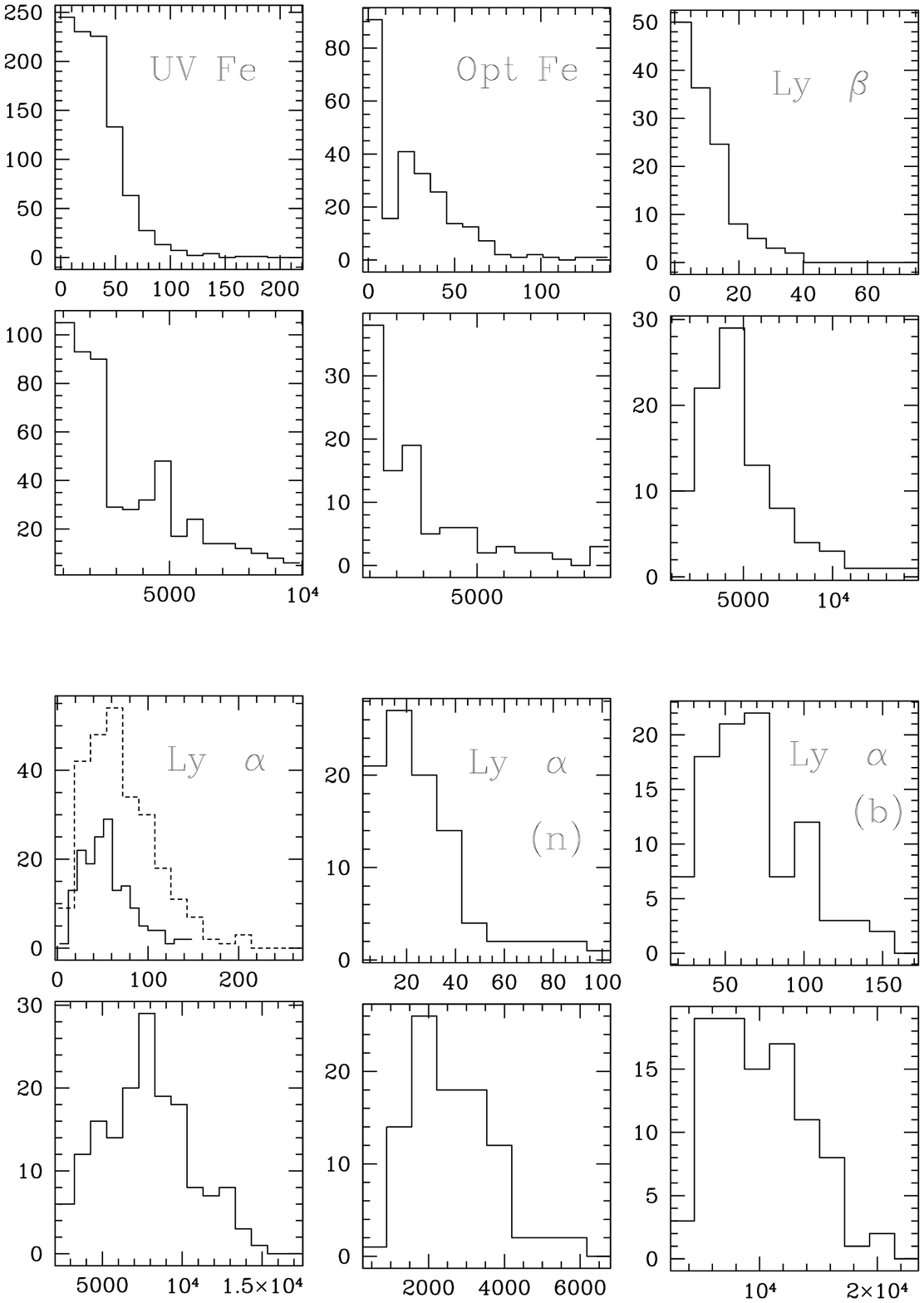}{15cm}{0}{50}{50}{-280}{-30}
\plotfiddle{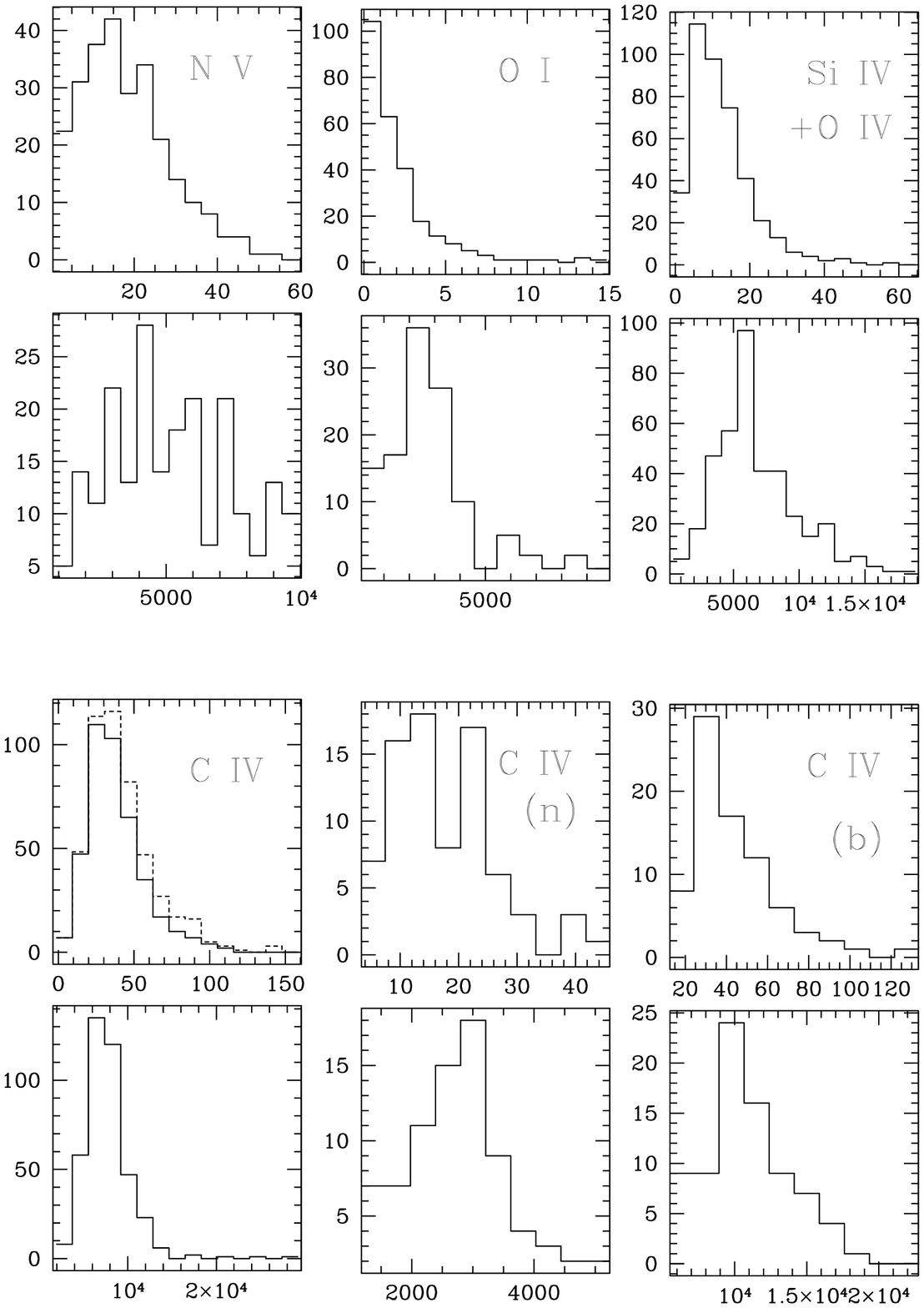}{1cm}{0}{50}{50}{-20}{10}
\caption{The KM estimated distribution of the emission line properties
of the LBQS sample. The upper panels are rest frame \ew~ (\AA) and the
lower panels are the FWHM (\kms) of the Gaussian profile used
to model the emission lines (see \S3.2 and \S4 for an explanation of the
meaning of the iron emission \ew~ and FWHM). For emission lines that
may be modeled with two Gaussian profiles, the narrow component
distributions are marked (n) and broad components (b). The
distribution of the sum of the narrow + broad component \ew~ is also
shown as the dashed line in the panels for the single component
models.}
\end{figure}

\clearpage
\eject

\begin{figure}[h]
\figurenum{4}
\plotfiddle{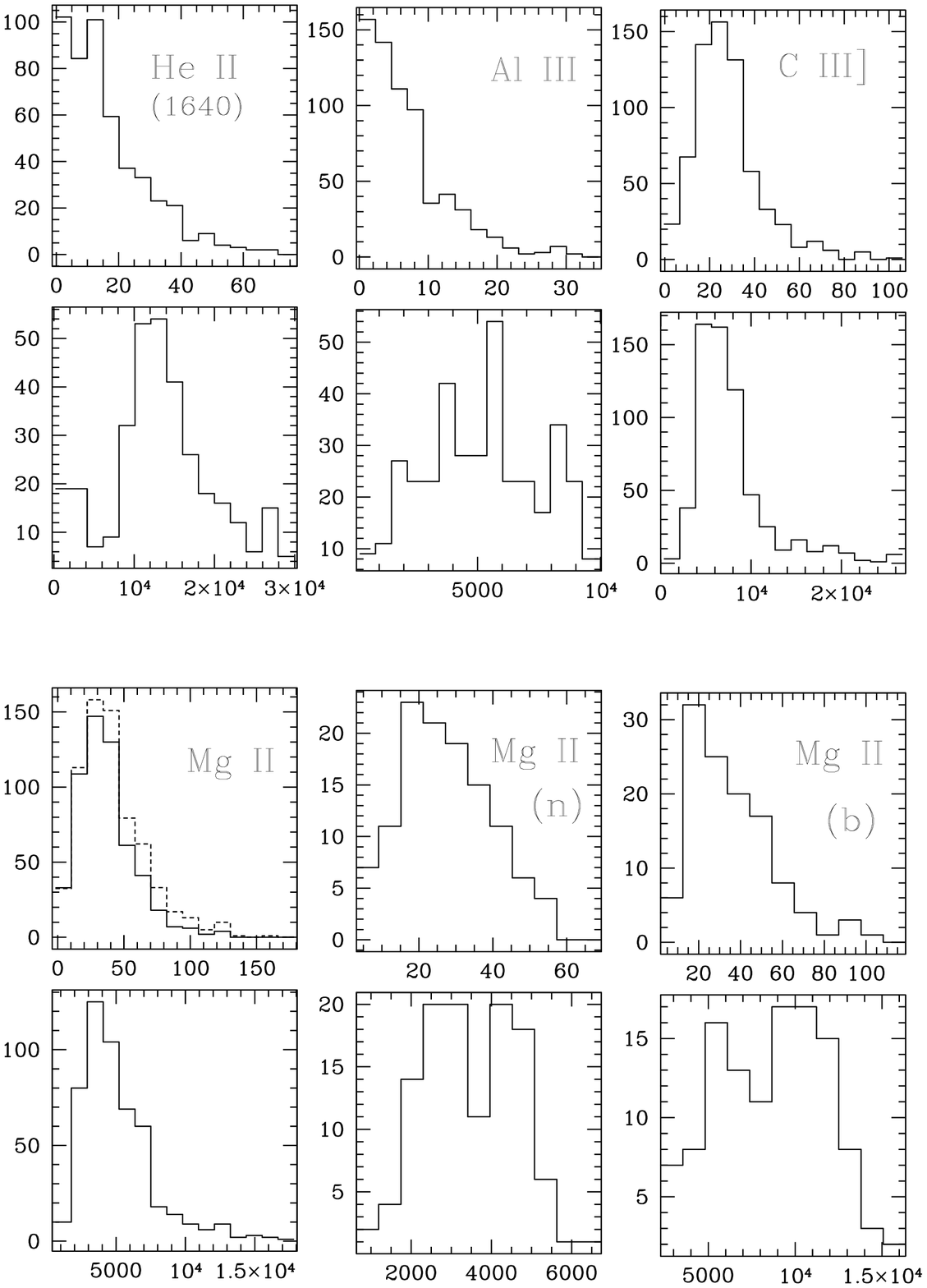}{15cm}{0}{50}{50}{-280}{-30}
\plotfiddle{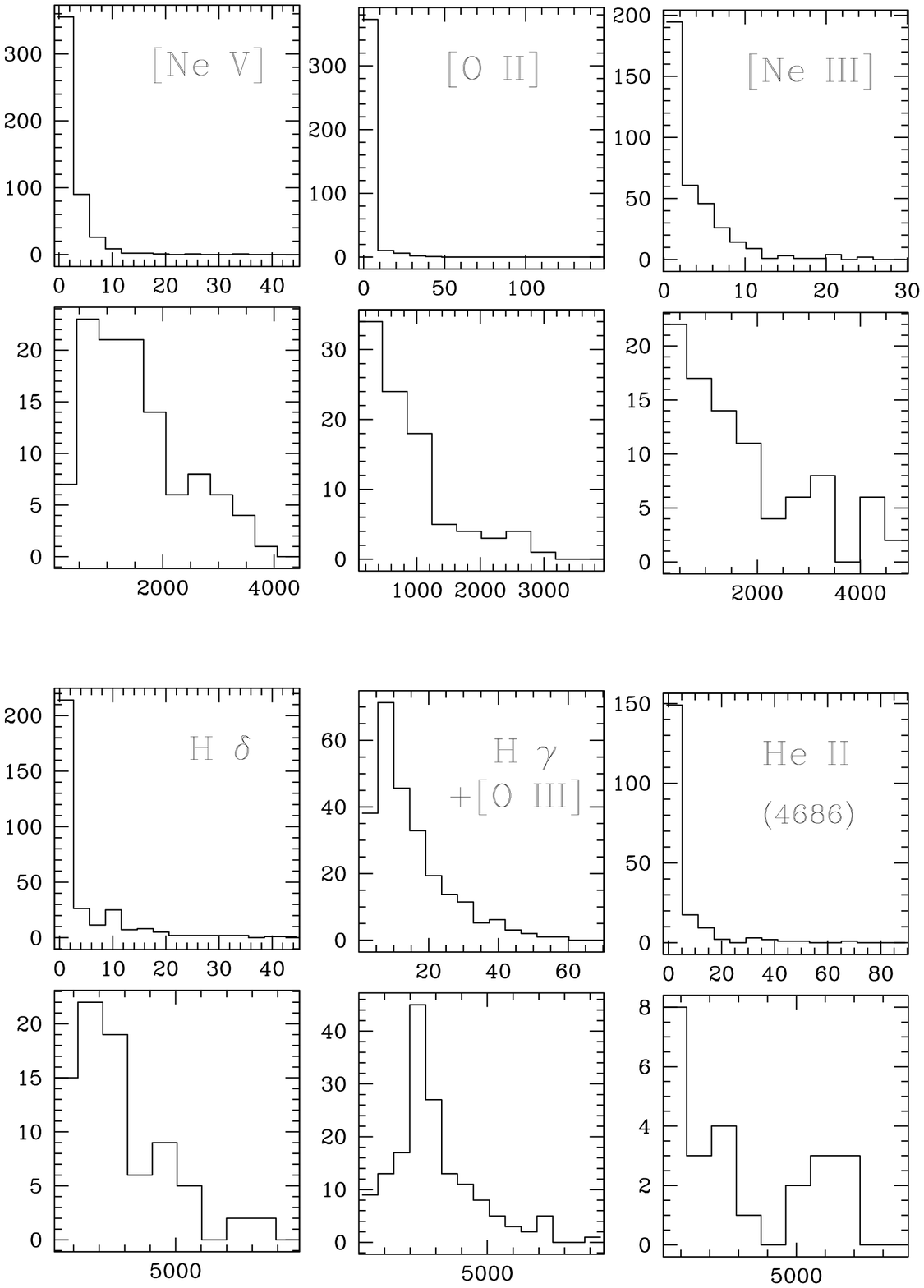}{1cm}{0}{50}{50}{-20}{10}
\caption{Cont.}
\end{figure}

\clearpage
\eject

\begin{figure}[h]
\figurenum{4}
\plotfiddle{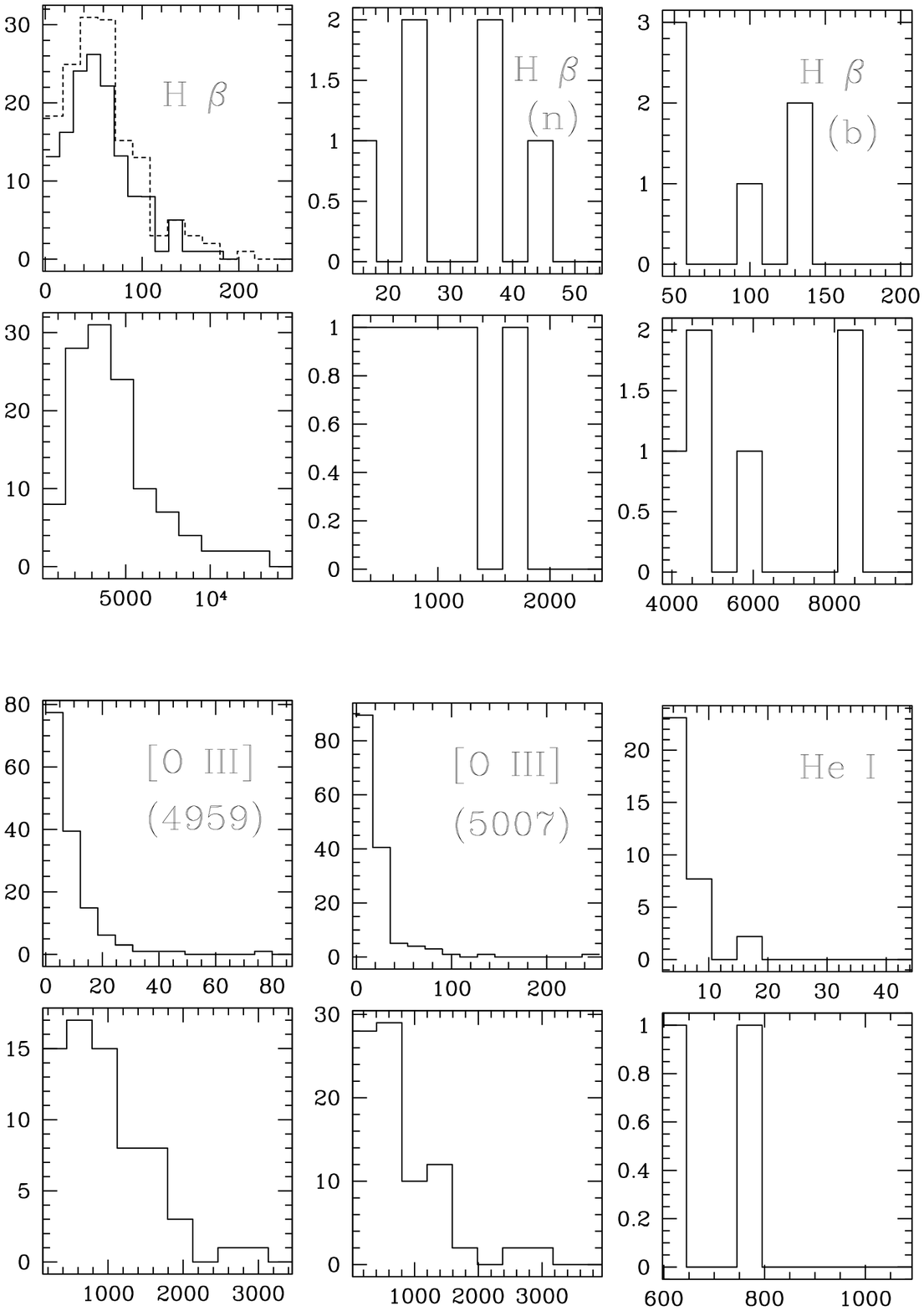}{15cm}{0}{50}{50}{-100}{0}
\caption{Cont.}
\end{figure}

\clearpage
\eject

\end{document}